\documentclass[pre,twocolumn]{revtex4-1}
\usepackage{ae}
\usepackage{amsmath}
\usepackage{amsfonts}
\usepackage{fontenc}
\usepackage[dvips]{graphicx}
\usepackage{fancyhdr}
\usepackage{color}
\usepackage{xcolor}
\usepackage{float}
\usepackage{hyperref}
\usepackage{tikz}

\newcommand{\medio}[1]{\left\langle #1\right\rangle}
\DeclareMathOperator{\sign}{sign}

\begin{document}

\title{Zealots in the mean--field noisy voter model}
\author{Nagi Khalil, Maxi San Miguel, and Raul Toral}
\affiliation{IFISC (CSIC-UIB), Instituto de Física Interdisciplinar y Sistemas Complejos, Campus Universitat de les Illes Balears, E-07122 Palma de Mallorca, Spain}

\begin{abstract}
The influence of zealots on the noisy voter model is studied theoretically and numerically at the mean--field level. The noisy voter model is a modification of the voter model that includes a second mechanism for transitions between states: apart from the original herding processes, voters may change their states because of an intrinsic, noisy in origin source. By increasing the importance of the noise with respect to the herding, the system  exhibits a finite--size phase transition from a quasi-consensus state, where most of the voters share the same opinion, to a one with coexistence. Upon introducing some zealots, or voters with fixed opinion, the latter scenario may change significantly. We unveil the new situations by carrying out a systematic numerical and analytical study of a fully connected network for voters, but allowing different voters to be directly influenced by different zealots. We show that this general system is equivalent to a system of voters without zealots, but with heterogeneous values of their parameters characterizing herding and noisy dynamics. We find excellent agreement between our analytical and numerical results. Noise and herding/zealotry acting together in the voter model yields not a trivial mixture of the scenarios with the two mechanisms acting alone: it represents a situation where the global--local (noise--herding) competitions is coupled to a symmetry breaking (zealots). In general, the zealotry enhances the effective noise of the system, which may destroy the original quasi--consensus state, and can introduce a bias towards the opinion of the majority of zealots, hence breaking the symmetry of the system and giving rise to new phases. In the most general case we find two different transitions: a discontinuous transition form an asymmetric bimodal phase to an extreme asymmetric phase and a second continuous transition from the extreme asymmetric phase to an asymmetric unimodal phase.
\end{abstract}

\date{\today}
\maketitle

\section{Introduction \label{sec:1}}

The voter model is a paradigmatic non-equilibrium system that has been used, amongst other applications, to study the evolution to consensus in a population \cite{clsu73,holi75,li05,castellano09,kaprisky10}. In the model, a set of connected individuals or agents, known generically as {\sl voters}, can switch between two opinion states by copying the state of a randomly chosen neighbor. In the simplest, mean--field, formulation all agents are identical and connected to all others, in such a way that a voter supporting a particular value of the opinion can change it with a rate proportional to the fraction of agents holding the opposite one, the proportionality constant named as the {\sl herding} parameter, $h$. If the system is finite, the model exhibits a competition that, after a transient time, ends up in the absorbing state of global consensus where all agents hold exactly the same opinion and no further evolution is possible. This picture may change when the model is modified in order to account for more realistic situations. Among all modifications studied in the literature, we consider heterogeneity and noise. For other aspects which are usually accounted for by an statistical physics approach to social dynamics see \cite{pejorawa17}.

Heterogeneity appears, for example, when individuals differ by their intrinsic rates of change between states, an extreme case being that of a {\sl zealot}, an agent that never changes his state. Another source of heterogeneity arises when some agents are able to copy only a subset of the whole population, a situation naturally described by a graph or network of interactions \cite{suchecki}. Previous work has focused on the effect that a few zealots might have on the asymptotic states for regular networks or all-to-all interactions \cite{mo03,mopere07,fues14,X15} and, more recently for complex networks \cite{kahoda14,moscag15,fues14,X15}. In these cases, the existence of zealots changes drastically the evolution of the system: if only one zealot is present, the system approaches much faster one absorbing state, the state of consensus that corresponds to the zealot; for equal number of zealots of different opinions, the system reaches a dynamically active non-consensus steady state. The presence of zealots in nonlinear voter models has been shown to result in a rich phenomenology \cite{mo15,memozi16,memozi17}. See also \cite{gamo91,ga97,gaja07,xisrkozhlisz11,ma12,librwashstha13,ma13,tuwegr13,veswch14,nama15,coca16,klwidodo17} for recent papers addressing the influence of zealots or other agents whose opinions have a special weight in the dynamical rules on a variety of models of cooperation and opinion dynamics, and in the influence of zealots in spatial rock-paper-scissors game \cite{szpe16}. Zealots are  frequently considered also in the realm of evolutionary games and  research concerning the evolution of cooperation, as reviewed recently  in \cite{pejorawa17}.

Noise has been included in the voter model as an intrinsic tendency to spontaneous changes of state. In the so--called noisy voter \cite{grma95} or Kirman model for financial markets \cite{ki93,alluwa05,alluwa08,almi09,almira13} the rate at which one voter changes opinion includes, besides the dependence on the fraction of neighbors in the opposite state, an intrinsic constant or {\sl noise parameter}, $a$. {This way of introducing noise can be easily adapted to different models of opinion dynamics, such as \cite{deneamwe00,wedeamna02,hekr02,wedeamna03}. In any case, the} main effect of noise is that there are no absorbing states so that it prevents the system from reaching the full consensus states. Moreover, by increasing the ratio $a/h$ of the noise to the herding parameter the system undergoes at a critical value $(a/h)_c$ a finite-size transition from a bimodal phase (where agents spend most of the time close to one consensus states and then switching to the other consensus) to a unimodal one (where there is coexistence of two macroscopic subpopulations at different states) \cite{catosa15,catosa16}. The presence of a complex network seems not to change this general picture, while the critical value $(a/h)_c$ is modified. Few studies consider agents heterogeneity in the context of the noisy voter model, see \cite{lato13} as an exception, nor the influence of zealots.

In this work, we analyze the effects of zealotry on the noisy voter model, focusing on the steady--state properties, and provide a deep relation between this system and a system made of heterogeneous voters. We restrict our study to the simplest case of all--to--all or mean--field interaction{, since it represents a suitable and simple scenario where the competition of global (noise) and local (herding/zealotry) mechanisms is coupled to a symmetry breaking induced by zealots, which can also be global or local. Apart from the latter general and physical motivation, we aim at understanding} the role played by zealots on a population of agents whose dynamics accounts for two important processes, namely copying or herding and intrinsic noise. More specifically, we want to describe, quantify, and understand the changes induced on the different phases of the noisy voter model. {In this way, the present study is a natural extension of previous works on opinion dynamics of voters \cite{fesurasaeg14}, the zealots representing now leaders or inflexible voters, for instance. We sustain our study on two complementary approaches: a theoretical one, based on a master--equation description, and a numerical one. For all the cases studied, the two approaches compare almost perfectly}.

The paper is organized as follows. In Sec. \ref{sec:2} we introduce the stochastic model of homogeneous noisy voters with different subgroups or communities affected by a different number of zealots. By an appropriate redefinition of the constants, we show that the model describes a set of heterogeneous noisy voters without zealotry, the effect of zealots being accounted for in the new rate constants. Sec.~\ref{sec:3} considers the simplest situation of one single community, and corresponds to a global influence of zealots since all individuals are equally influenced. {The latter abstract scenario can be shown as an homogeneous population of voters equally affected by one or several leaders}. Despite its simplicity, this case turns out to be important because it is tractable analytically. Moreover, it provides available information for the study of more general cases, in particular the two community case considered in Sec. \ref{sec:4}. This two-community case, {seen now as different leaders acting on two different subset of an homogeneous population of voters}, is the minimal situation where the system can exhibit all possible phases, and a suitable context to compare the approximate theories, discussed in Sec. \ref{sec:4}, Sec. \ref{sec:5}, and Appendix \ref{app:1}, against the numerical simulations. {That way the theory is constructed going from simple and concrete cases to general ones, gaining step by step understanding}. Two complementary theories are provided in this work. One more general but approximated, based on the analysis of the master equation, given in Sec. \ref{sec:5}; and another one which is exact but restricted to the case of one community, at Appendix \ref{app:1}. Finally, Sec. \ref{sec:6} is devoted to the {discussion and} conclusions.

\section{Model \label{sec:2}}
\begin{figure}[!h]
 \centering
 \includegraphics[width=.3\textwidth]{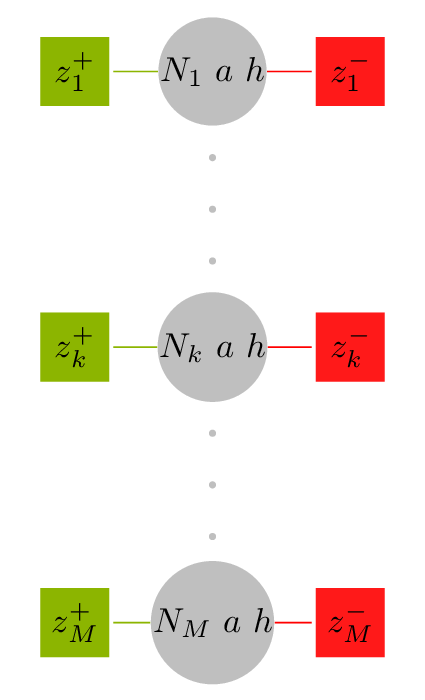}
 \caption{Schematic representation of the system made of agents having the same constants $a$ and $h$ but divided into $M$ communities with different sets of voters directly influenced by different zealots. The same color of circles (communities) indicates common constants, a community being fully characterized by the number of optimistic (green--left square) and pessimistic (red--right square) zealots linked to it. {For an alternative representation showing the fine structure of one zoomed community see Fig.\ref{fig:3}}.}
 \label{fig:1}
\end{figure}

We consider $N$ agents, each one capable to be in one of two possible states. Following the original application to financial markets by Kirman \cite{ki93}, we name those states as ``optimistic'' and ``pessimistic'' but in this work we do not give any particular meaning to the states of the agents. The system is divided into $M$ subsystems or communities so that community $k$ has $N_k$ agents, $n_k$ of them being optimistic at a given moment, under the influence of $z_k^+$ optimistic and $z_k^-$ pessimistic zealots, as schematically represented in Fig.~\ref{fig:1}. We stress, however, that each agent interacts with any other agent, irrespectively of the community they belong to.

Since all agents are identical inside their communities, the state of the system is specified by the set $S=\{n_1,\dots,n_M\}$ of the number of optimistic agents of each subsystem. The dynamics or time evolution of $S$ is given by a stochastic process characterized by the rates $\pi^{+}_k$ and $\pi^{-}_k$ for transitions $n_k\to n_k+1$ and $n_k\to n_k-1$, respectively:
\begin{equation}
 \label{eq:1}
 \begin{split}
 & \pi^{+}_k=\left(a+h\frac{n+z_k^+}{N+z_k}\right)(N_k-n_k), \\
 &\pi^{-}_k=\left(a+h\frac{N-n+z_k^-}{N+z_k}\right)n_k,
 \end{split}
\end{equation}
where $z_k=z_k^++z_k^-$ is the total number of zealots of community $k$ and $n=\sum_kn_k$ is the total number of optimistic agents. Observe that the interaction among agents and among agents and zealots is different. While the former involves all possible pairs of agents, regardless the community they belong to, the latter distinguishes between communities. As described in the introduction, the rates have two contributions: the one encoded by the noise constant $a\ge 0$ is such that a voter changes his opinion randomly regardless the opinions of other voters or zealots; the contribution encoded by the herding constant $h\ge 0$ represents the random copying mechanism whose rate is proportional to the total number of voters with opposite opinion in the whole system, plus the number of zealots with different opinion within the same community. We recover the usual noisy voter model if $z_k=0$, and the voter model if, in addition, $a=0$.
 
\begin{figure}[!h]
 \centering
 \includegraphics[width=.23\textwidth]{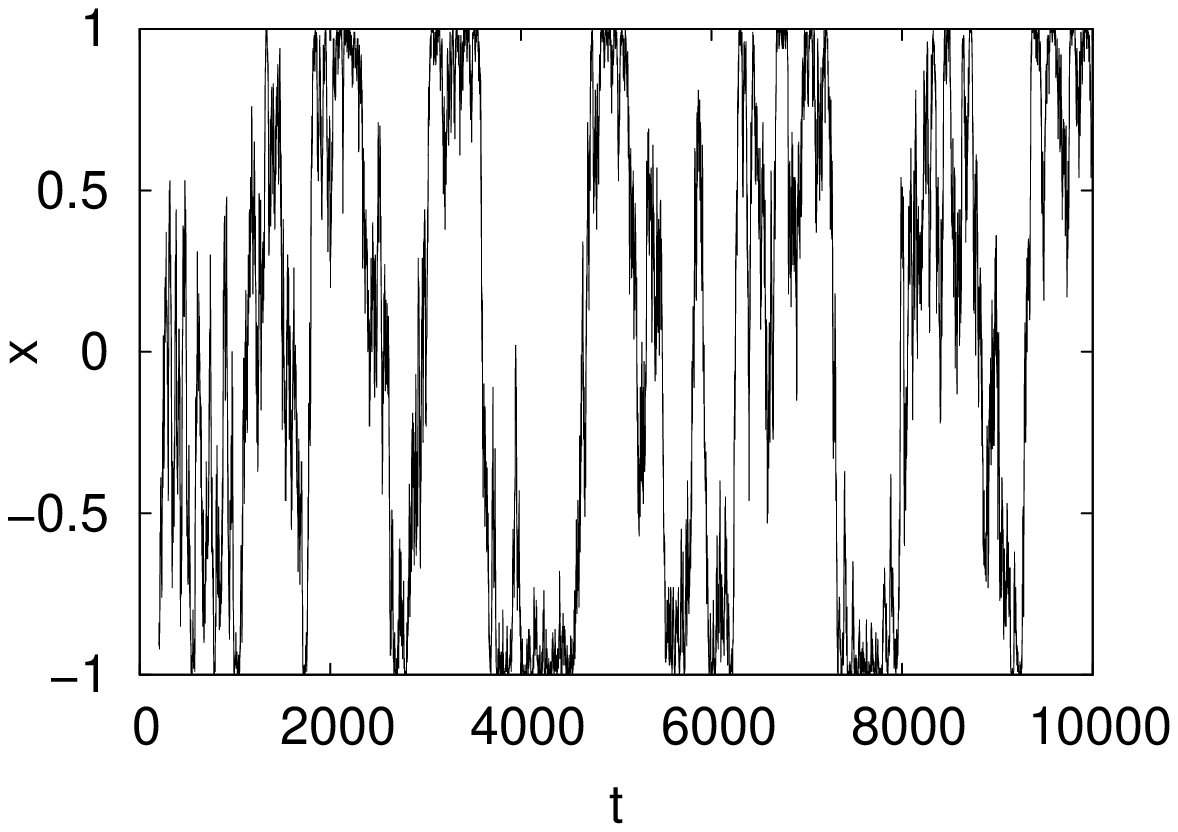}
 \includegraphics[width=.23\textwidth]{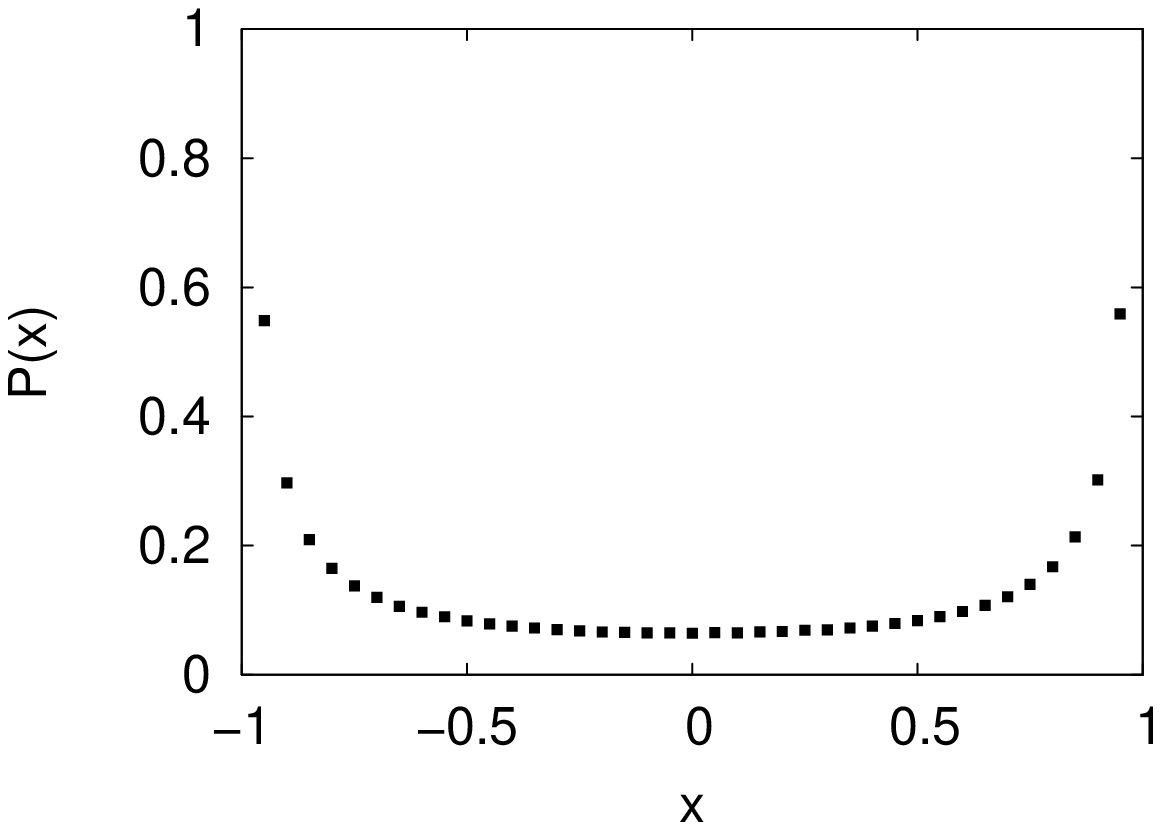}
 \includegraphics[width=.23\textwidth]{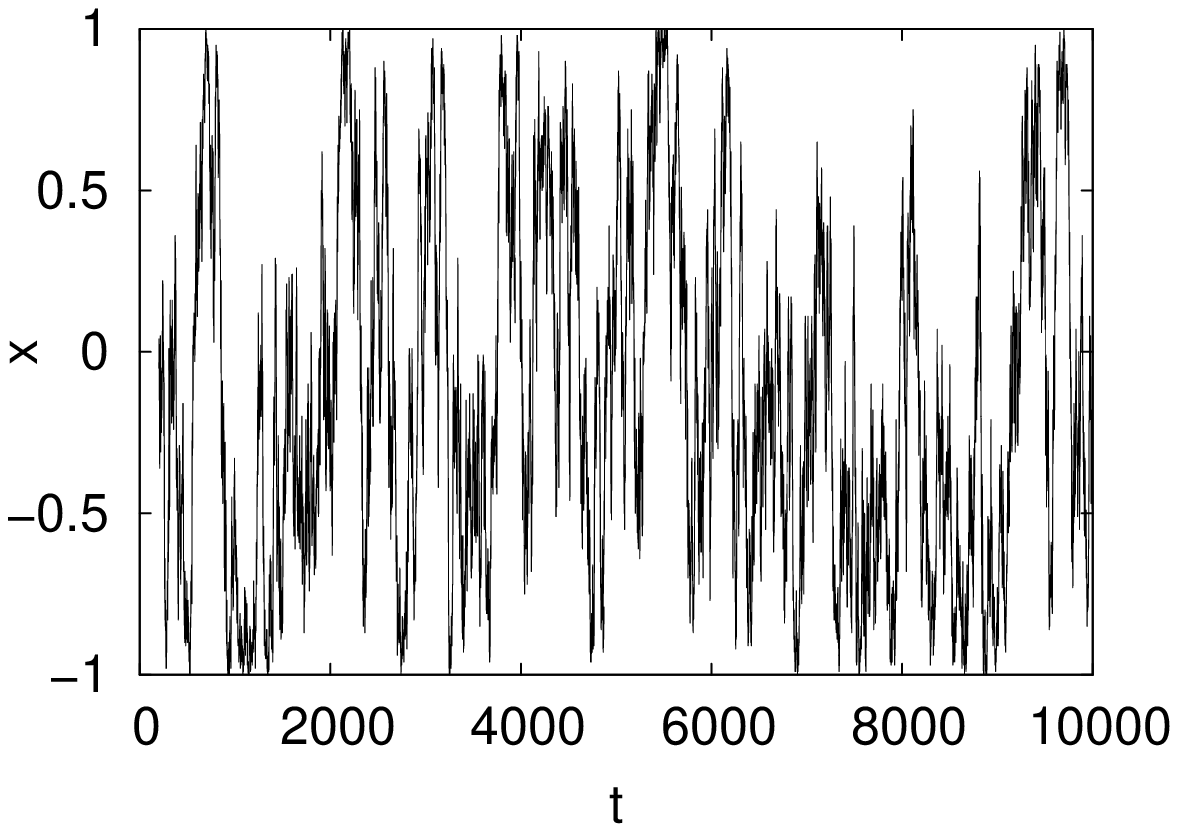}
 \includegraphics[width=.23\textwidth]{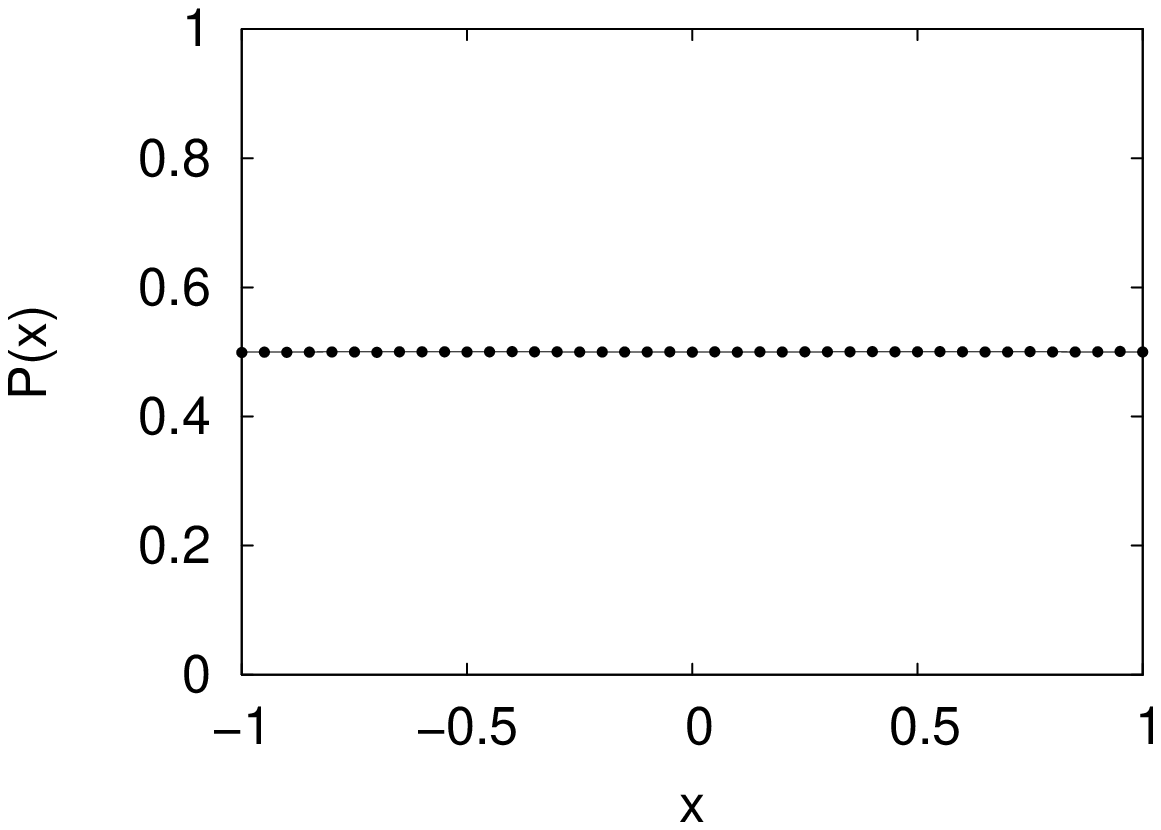}
 \includegraphics[width=.23\textwidth]{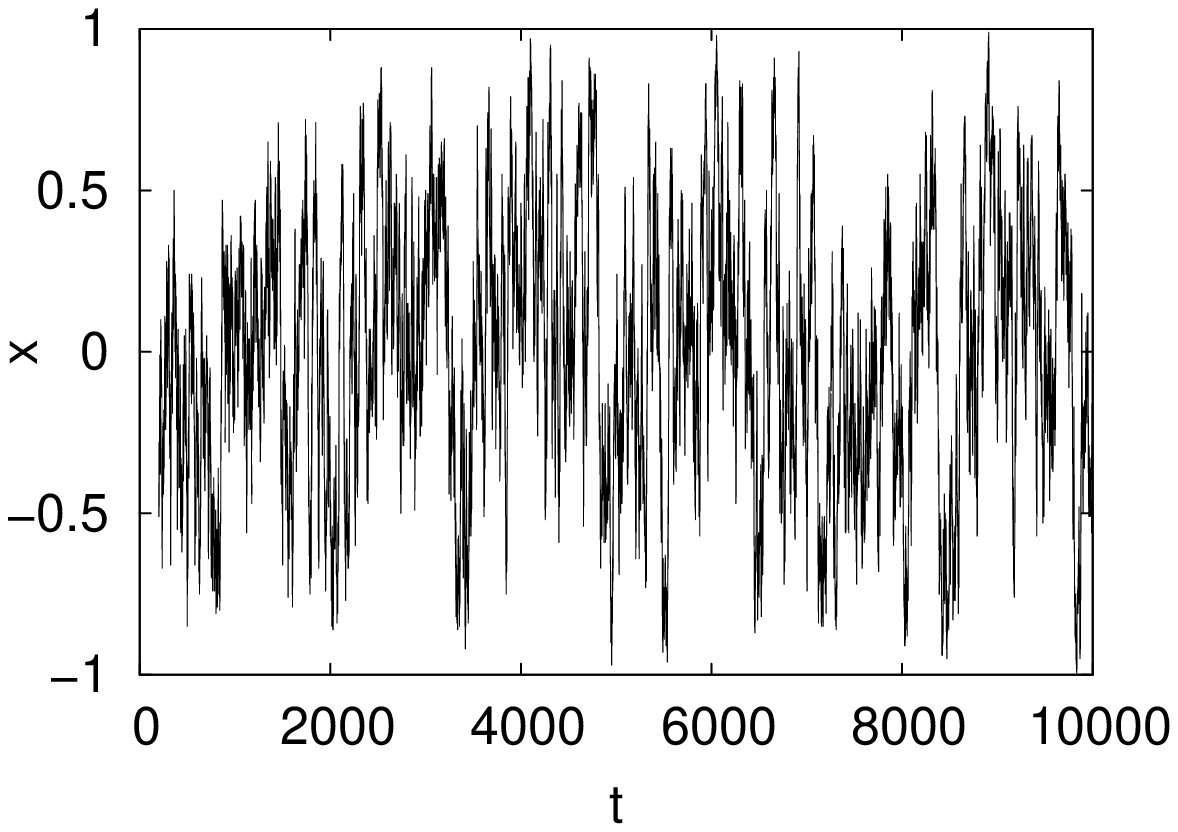}
 \includegraphics[width=.23\textwidth]{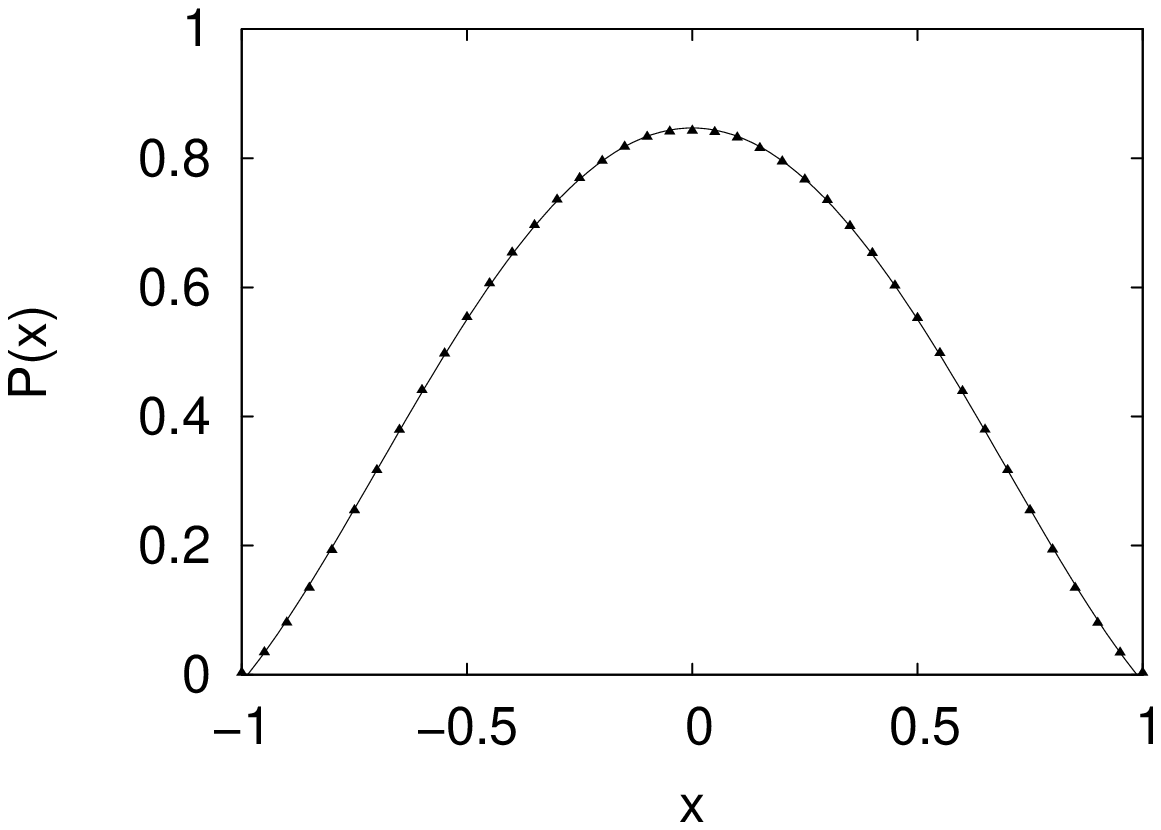}
 \caption{Simulation results for the trajectory $x(t)$ and probability function $P(x)$ of a system of $N=200$ agents, $N_1=N/2$, $z_1^+=z_1^-=1$, and $z_2=0$
for $a/h\ll (a/h)_c$ (plots on the top), $a/h=(a/h)_c$ (center), and $a/h\gg (a/h)_c$ (bottom). Time is measured in units of $h^{-1}$. The lines on the plots of $P(x)$ are the reconstruction of the probabilities using $K=4$ moments as explained in Appendix \ref{app:2}.}
 \label{fig:8}
\end{figure}

From a mesoscopic point of view, the system is characterized by the probability $p(S)$ of the system being in a state $S$. The master equation for $p(S)$, and its corresponding moments, are easily obtained from rates \eqref{eq:1} or \eqref{eq:2}. As discussed in the Appendix \ref{app:3}, the equations for the dynamical evolution of the moments of a given order involve only moments of lower order and, hence, are closed. For the usual noisy voter or Kirman model (with no zealots) the analysis of the stochastic system proves that there is a finite size phase transition characterized by a qualitative change of the steady-state probability distribution $P(x)$ of observing a ``magnetization'' $x=2\frac{n}{N}-1$. Fig.~\ref{fig:8} shows typical trajectories $x(t)$ and their respective steady probability functions $P(x)$. As it is apparent from this figure, the system may exhibit two phases separated by a critical value $(a/h)_c=1/N$. For $a/h\le (a/h)_c$ the system is in the {\slshape symmetrical bimodal} (SB) phase where voters share the same opinion most of the time, having the two opinions the same overall probabilities in the long run, hence a typical trajectory has long stays with extreme values of the magnetization and short transitions among them, while the steady probability function $P(x)$ accumulates around the extremes and becomes bimodal with symmetric maxima at $x_m=\pm1$. For $a/h\ge (a/h)_c$ the system is in the {\slshape symmetric unimodal} (SU) phase where probabilities accumulate around $x_m=0$, corresponding to a coexistence of opinions. In the border case $a/h=(a/h)_c$ the probability function $P(x)$ is uniform in the $x\in[-1,1]$ space, meaning that any fraction of optimistic voters are equally probable. See also plot (a) of Fig.~\ref{fig:4}. It is our objective in this paper to investigate the effect that the presence of zealots and the splitting in communities has on the unimodal-bimodal transition.

\begin{figure}[!h]
 \centering
 \includegraphics[width=.25\textwidth]{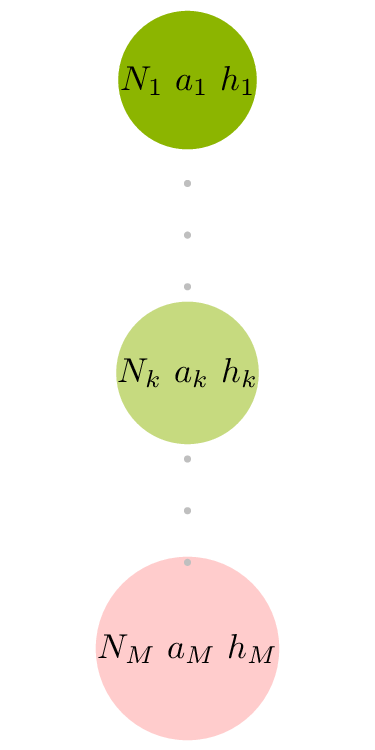}
 \caption{Schematic representation of the system where no zealots is present but different communities have different constants (colors--positions).}
 \label{fig:2}
\end{figure}

It is worth writing the rates of Eq.~\eqref{eq:1} as
\begin{equation}
 \label{eq:2}
 \begin{split}
 \pi^{+}_k&= \left(a_k^++h_k\frac{n}{N}\right)(N_k-n_k), \\
 \pi^{-}_k&=\left(a_k^-+h_k\frac{N-n}{N}\right)n_k,
 \end{split}
\end{equation}
with
\begin{equation}
 \label{eq:3}
 \begin{split}
 & a_k^\pm\equiv a+\frac{z_k}{2N}h_k\pm \frac{\Delta z_k}{2N}h_k,\quad
 \Delta z_k\equiv z_k^+-z_k^- \\
 & h_k\equiv \frac{N}{N+z_k}h,.
 \end{split}
\end{equation}
In this way we show that the system of $N$ identical noisy voters (same $a$ and $h$), under the influence of zealots, is equivalent to $N$ noisy voters without any zealotry influence but with some heterogeneity in the noise and herding constants, see Fig.~\ref{fig:2}. If $\Delta z_k\ne 0$, the original noise parameter $a$ splits in two: $a^+_k$ for pessimistic to optimistic transitions, and $a_k^-$ for the optimistic to pessimistic ones, that is, inducing a bias in between the two states. On the contrary, the herding parameter is not affected in this sense, being the same for both transitions. In all cases, the mean noise parameter $(a_k^++a_k^-)/2$ increases with the total number of zealots acting on community $k$, while the herding parameter $h_k$ decreases.

\section{Global influence \label{sec:3}}
In this section we consider a single community, $M=1$, under a global influence of zealots. It is convenient to distinguish between balanced (equal number of pessimistic and optimistic zealots) and unbalanced cases.

\subsection{Balanced case}

\begin{figure}[!h]
 \centering
 \includegraphics[width=.4\textwidth]{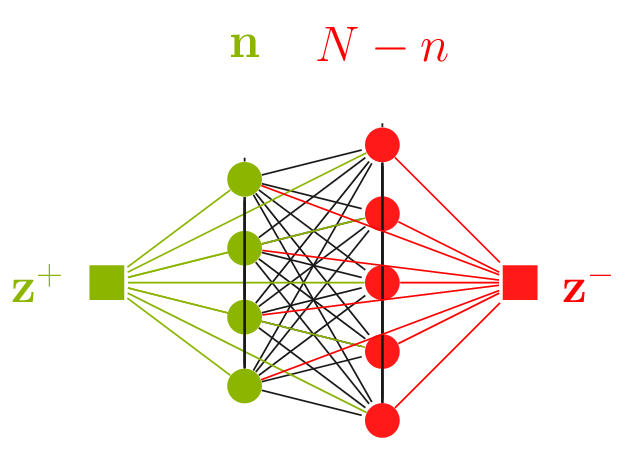}
 \caption{Schematic representation of the state of a system at a given time with one community of optimistic (green--left circles) and pessimistic (red--right circles) agents, influenced by $z^+$ optimistic (green--left square) and $z^-$ (red--right square) pessimistic zealots.}
\label{fig:3}
\end{figure}

Consider a situation like in Fig.~\ref{fig:3}, with one community $M=1$ and the same number of optimistic and pessimistic zealots $z^+=z^-=z/2$.

From Eq.~\eqref{eq:2}, we obtain that the system behaves like a noisy voter model without zealots with effective noise and herding parameters
\begin{equation}
 \label{eq:6}
 \begin{split}
 &a_b=a+\frac{z}{2(N+z)}h, \\
 &h_b=\frac{N}{N+z}h,
 \end{split}
\end{equation}
where $a_b$ increases with $z$, while $h_b$ decreases (the subscript ``$b$'' refers to the balanced case). That is to say, the net effect of zealotry is to enhance the original noise by increasing $a_b/h_b$.

The functional form of the new constants at Eq.~\eqref{eq:6} are easily understood if we look at the dynamics at the agent level. The factor in $h_b$ is a direct consequence of removing the zealots from the system in the interpretation of Eq.~\eqref{eq:6}: an agent now can copy the opinion of only $N$ agents, while initially there where $N+z$ agents and zealots. The noise term accounts for the removal of zealots. The additional contribution to $a_b$ is essentially the rate at which one zealot was initially copied divided by two. The division by two is required since a couple of opposite zealots forms a unit of uncertainty, or equivalently because only half of the zealots contribute to either one of the two possible transitions. This picture clarifies the deep connection between the voter and the noisy voter model, in the sense that the latter can be understood as the former with the additional influence of couples of opposite zealots.

\begin{figure}[!h]
 \centering
 \includegraphics[width=.5\textwidth]{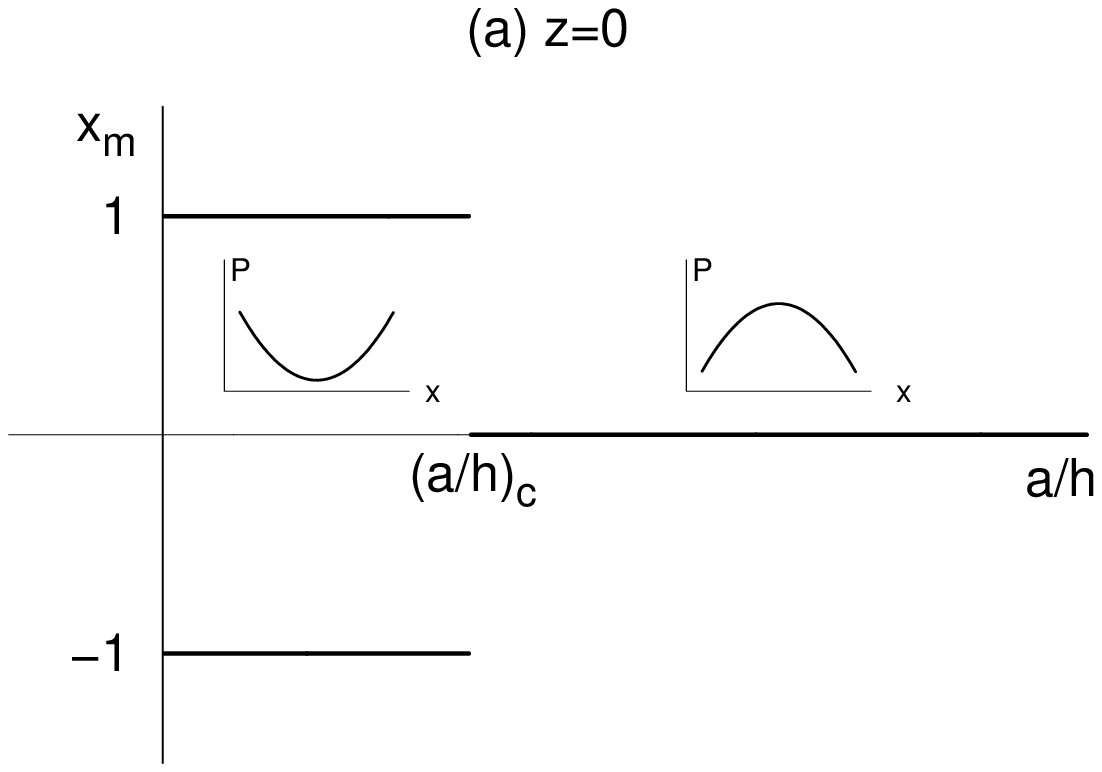} 
 \includegraphics[width=.5\textwidth]{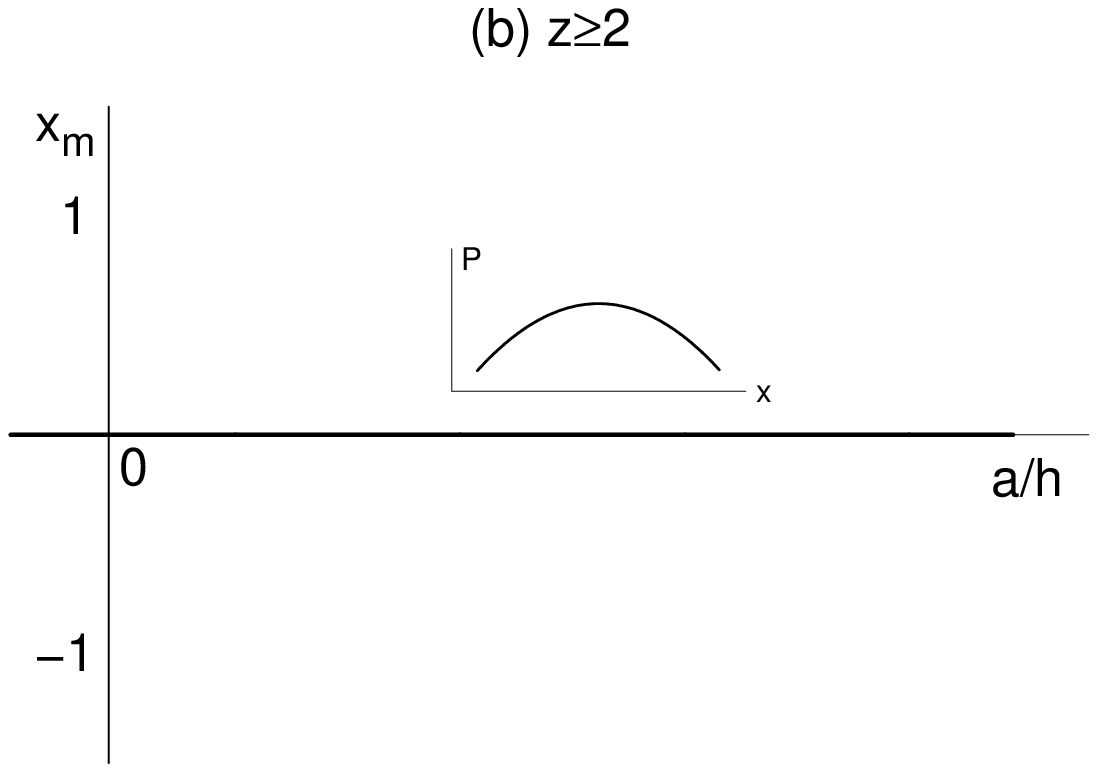} 
 \includegraphics[width=.475\textwidth]{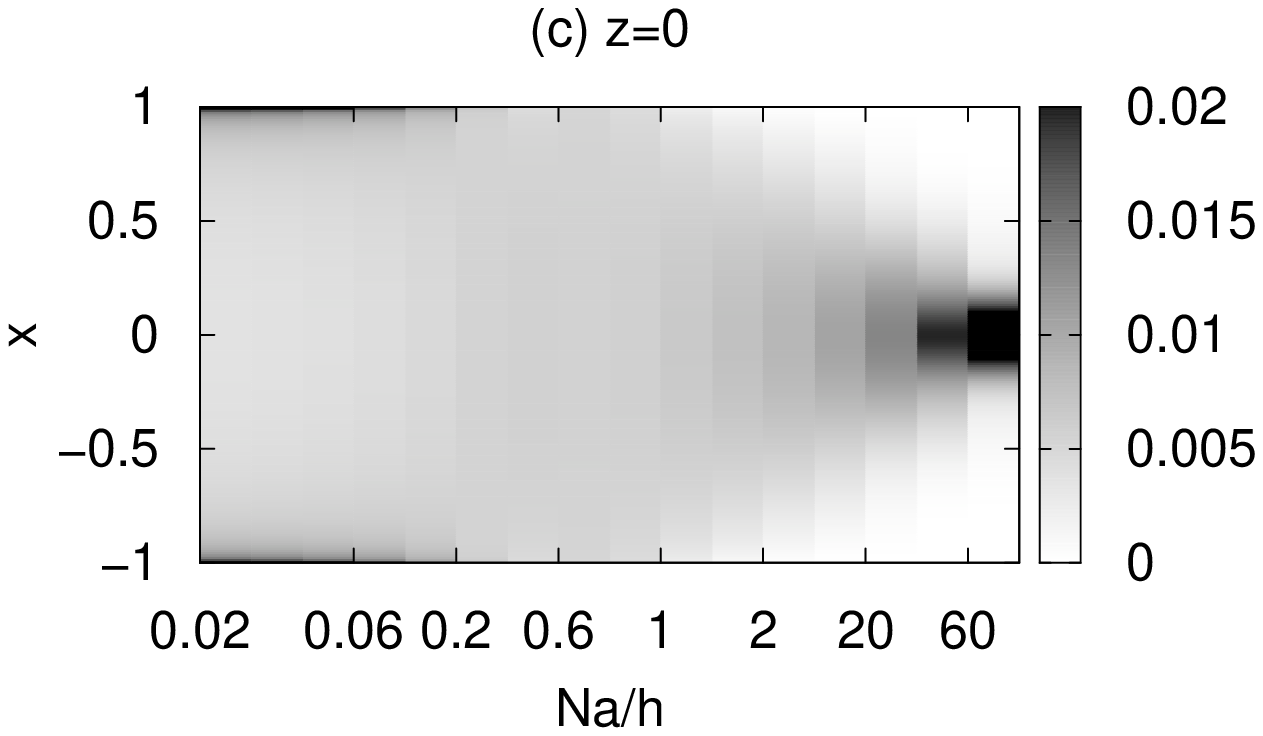}
 \caption{{Plots (a) and (b):} position of the maxima of $P(x)$ and its different phases as a function of $a/h$ when (a) $(a/h)_c>0$ and (b) $(a/h)_c\le0$. The inserts show schematically the shape of the probability distribution $P(x)$: bimodal for $(a/h)<(a/h)_c$ and unimodal for $(a/h)>(a/h)_c$. {Plot (c): probability distribution (gray scale) as a function of $x$ (vertical axis), for different values of $a/h$ (horizontal axis).}}
 \label{fig:4}
\end{figure}

As for the noisy voter model, the border case separating symmetric bimodal and symmetric unimodal is given by the condition $a_b/h_b=1/N$, that using Eq.~\eqref{eq:6} reads
\begin{equation}
 \label{eq:7}
 (a/h)_c=\frac{2-z}{2(N+z)}.
\end{equation}
Since we are considering $z\ge 2$ (remember that for the present case $z$ is an even number), the critical value given by Eq.~\eqref{eq:7} is zero or negative. That means that for $z^+=z^-\ge 1$ the system only shows the symmetric unimodal phase. In other words, zealotry always destroys the symmetric bimodal phase, a result that also holds for $a=0$, the noiseless voter model, see plot (b) of Eq.~\ref{fig:4}. We realize now how sensitive is the system to the global influence of zealots: not only the absorbing or consensus states at $x=\pm 1$ disappear, but the most probable configuration of the system becomes one where equal fractions of agents with different opinions coexist. The dramatic change in passing from $z=0$ to $z=2$ is due to the global influence of zealots, and will be relaxed in Sec.~\ref{sec:4}, upon considering partial influence with two communities.

Concerning $x_m$, the location of the maxima of $P(x)$, we obtain $|x_m|=1$ if $a_b/h_b<1/N$ and $x_m=0$ if $a_b/h_b>1/N$, provided $z=0$; while it is $x_m=0$ for $z\ge 2$. As depicted in Fig.~\ref{fig:4}, $x_m$ experiences a discontinuous transition at the critical point Eq.~\eqref{eq:7} for $z=0$ (a), while the transition disappears for $z\ge 2$ (b). 

\subsection{Unbalanced case}
\begin{figure}[!h]
 \centering
 \includegraphics[width=.4\textwidth]{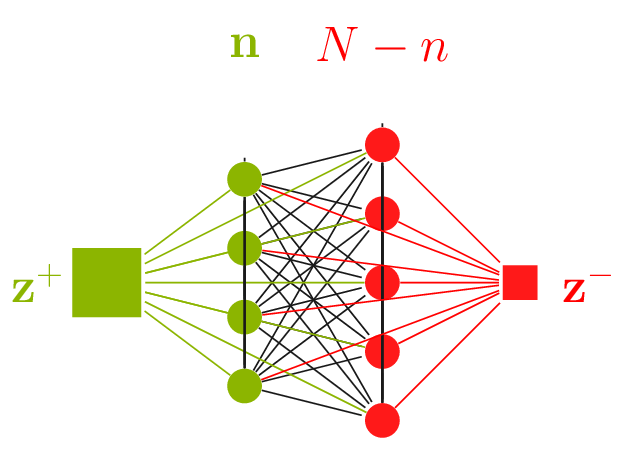}
 \caption{Schematic representation of the system with one community of optimistic (green--left circles) and pessimistic (red--right circles) agents, influenced by $z^+$ optimistic (green--left square) and $z^-$ (red--right square) pessimistic zealots. The different sizes of the squares indicates the different number of opposite zealots.}
\label{fig:5}
\end{figure}

We consider in this subsection $M=1$ and $z^+\ne z^-$, a situation schematically represented by Fig.~\ref{fig:5}.

Now, the dynamics of the system can be seen as that of a noisy voter model with two effective noise coefficients, $a^+_u$ for pessimistic to optimistic transitions, and $a^-_u$ for the reciprocal ones, and an effective herding constant $h_u$ (the subscript ``$u$'' refers to the unbalanced case). If the total number of zealots is $z$ and $\Delta z=z^+-z^-$, the new effective noise and herding parameters can be written as
\begin{equation}
 \label{eq:8}
 \begin{split}
 a_u^\pm&=a_b\pm \frac{\Delta z}{2N}h_u, \\
 h_u&=h_b.
 \end{split}
\end{equation}
with $h_b$ and $a_b$ given by Eq.~\eqref{eq:6}. Hence, the symmetry breaking induced by the zealots involves the noise contribution to the dynamics, the herding being modified by the total number of zealots $z$ regardless their preferred opinions. The form of the new coefficients can be interpreted with an agent-base analysis of the dynamics, as we did in the balanced case.

\begin{figure}[!h]
 \centering
 \includegraphics[width=.5\textwidth]{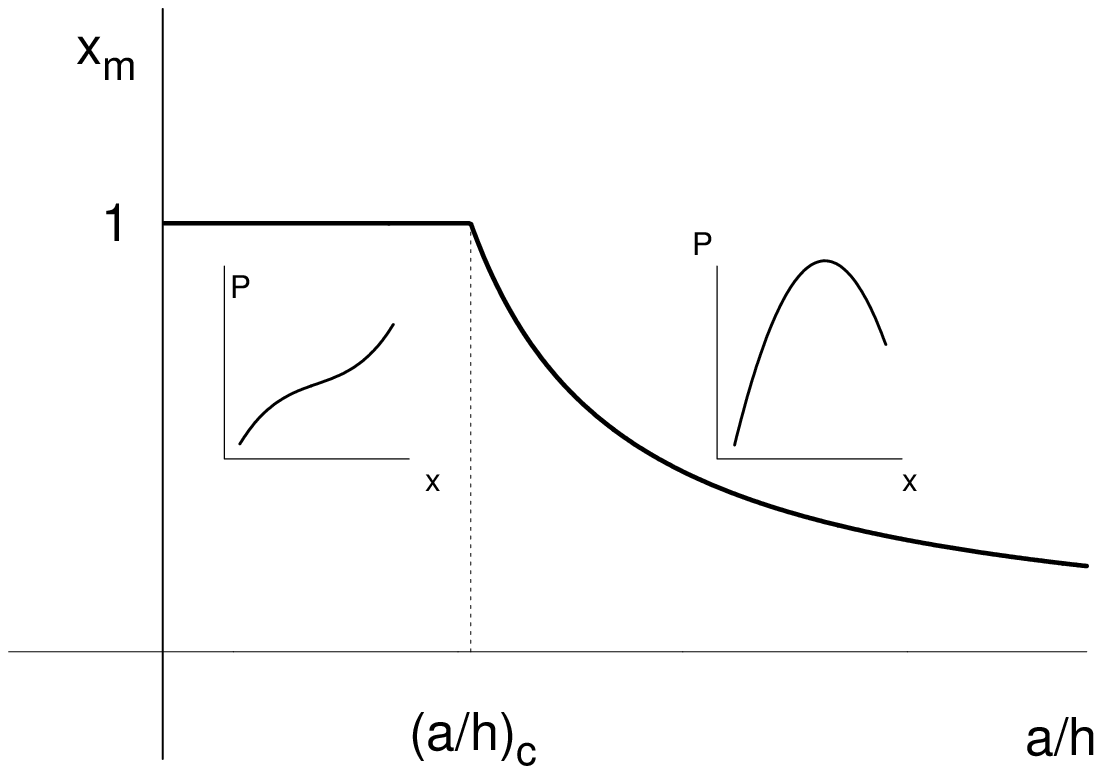} 
 \includegraphics[width=.475\textwidth]{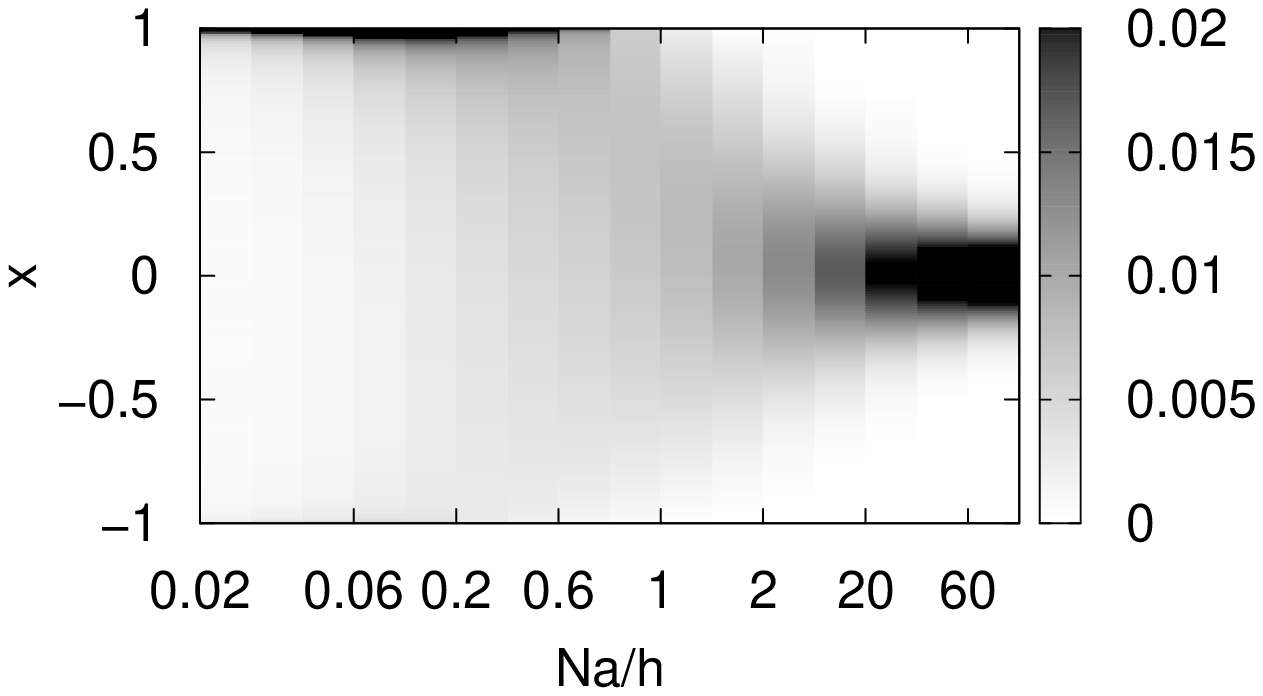}
 \caption{Top: position of the maxima of $P(x)$ and its different phases as a function of $a/h$ when $(a/h)_c>0$ and $\Delta z>0$. The inserts show schematically the shape of the probability distribution $P(x)$: extreme asymmetric (EA) for $(a/h)<(a/h)_c$ and asymmetric unimodal (AU) for $(a/h)>(a/h)_c$. {Bottom: probability distribution (gray scale) as a function of $x$ (vertical axis), for different values of $a/h$ (horizontal axis).}}
 \label{fig:6}
\end{figure}

As can be inferred from a study of a more general case to be carried out in Sec. \ref{sec:5}, and also in a different way in Appendix \ref{app:1}, the system can present two new phases (see Fig.~\ref{fig:6}). If the noise-herding ratio is smaller than the critical value
\begin{equation}
 \label{eq:9}
 (a/h)_c=\frac{1}{2(N+z)}\left(2-z+\frac{N+1}{N-1}|\Delta z|\right),
\end{equation}
the system is in the {\slshape extreme asymmetric} (EA) phase characterized by $P(x)$ having one minimum and one maximum at the extreme values of the magnetization (at $-\sign(\Delta z)$ and $\sign(\Delta z)$, respectively). For $a/h\ge (a/h)_c$, the absolute maximum of the previous phase moves to intermediate values of the magnetization and a local minimum appears at $x=\sign(\Delta z)$, the system being now at the {\slshape asymmetric unimodal} (AU) phase, where $P(x)$ displays a single maximum at $x_m\ne 0$.

The critical value given by Eq.~\eqref{eq:9} becomes negative when $z>2\frac{N+1}{N-1}+|\Delta z|\approx2+|\Delta z|$. In this case, the system exhibits always the AU phase, regardless the value of $a/h$. As in the balanced case, then, the zealotry destroys the preference for the consensus states favoring configurations where macroscopic fractions of agents with different opinions coexist. As we will see in Sec.~\ref{sec:4}, if zealots do not act upon all agents, new phases will appear for small values of $a/h$.

If we focus on the behavior of the maximum of the distribution, $x_m$, it is $x_m=\sign(\Delta z)$ in the extreme asymmetric phase, while it changes continuously as a function of $a/h$ in the asymmetric unimodal phase as
\begin{equation}
 \label{eq:10}
 x_m=\frac{\frac{N+1}{N-1}\Delta z}{2(N+z)\left[a/h-(a/h)_c\right]+\frac{N+1}{N-1}|\Delta z|},
\end{equation}
see Appendix \ref{app:1}. Contrary to the balanced case, now $x_m$ changes continuously in the transition, as shown in Fig.~\ref{fig:6}. The breaking of symmetry due to the unbalance in the number of optimistic and pessimistic zealots transforms the discontinuous behavior of $x_m$ into a continuous one.

\section{Partial influence \label{sec:4}}
Here we generalize the study of the previous section by allowing the zealots to directly influence only part of the system. In the general setup introduced in Sec.~ \ref{sec:2} we need to consider a system made of two communities: one with $N_1$ agents directly connected to the zealots and another with $N-N_1$ agents with no connections to zealots. The main objective is to describe the behavior of the voters as a whole allowing $N_1$ to vary from $0$ to $N$ and distinguishing again between balanced and unbalanced number of optimistic and pessimistic zealots.

\subsection{Balanced case}\label{sub:balanced}
\begin{figure}[!h]
 \centering
 \includegraphics[width=.4\textwidth]{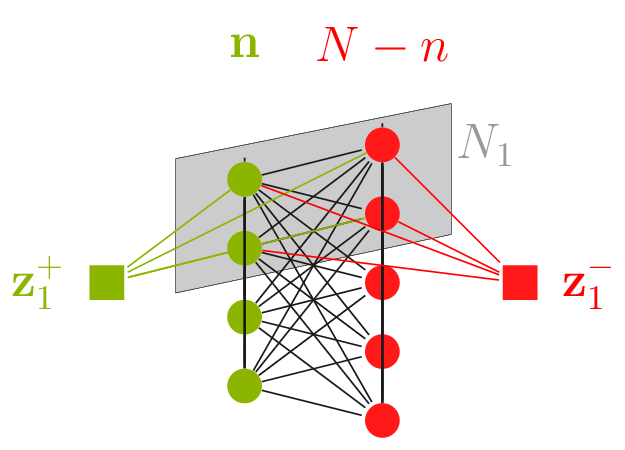}
 \caption{Schematic representation of the system of $M=2$ communities, only one directly influenced by the same number of optimistic and pessimistic zealots.}
\label{fig:7}
\end{figure}

We consider $M=2$ with $z_1^+=z_1^-$ and $z_2^+=z_2^-=0$, as schematized in Fig.~\ref{fig:7}.

Numerical simulations of this case unveil a similar phenomenology to that depicted in Fig.~\ref{fig:8}. The system exhibits two phases separated by a critical value $(a/h)_c$, that now depends on the number of zealots $z_1=z_1^++z_1^-$, $N$, and $N_1$. For $a/h\le (a/h)_c$ the system is in the symmetric bimodal while for $a/h\ge (a/h)_c$ the system is in the symmetric unimodal phase. Regarding the value of the maximum of the distribution, $x_m$, the situation is completely analogous to that of the one community case, as depicted in Fig.~\ref{fig:4}.

In order to derive an expression for $(a/h)_c$ in this case we can proceed as in the previous sections by trying to figure out how the dynamics of a single agent changes if we were to absorb the entire effect of the zealots into global noise and herding constants, now denoted by $a_b$ and $h_b$. In doing so, we approximate the dynamics by an effective one where all agents are equivalent. Consider first the herding constant. Upon eliminating the zealots, a fraction of $N_1/N$ copying processes are eliminated, hence $h$ is reduced by $\frac{N_1z_1}{N(N+z_1)}h$. The previous elimination produces a modification of the effective noise: we have to add to the parameter $a$ the contribution of the zealots, now $\frac{N_1z_1}{2N(N+z_1)}h$ for each possible transition. Hence, we have
\begin{equation}
 \label{eq:11}
 \begin{split}
 & a_b\simeq a+\frac{N_1z_1}{2N(N+z_1)}h, \\
 & h_b\simeq h-\frac{N_1z_1}{N(N+z_1)}h,
 \end{split}
\end{equation}
which in fact would be exact expressions if the two communities had the same statistical properties. Observe that the new constants are a generalization of Eqs. \eqref{eq:6} to situations of two communities. From Eqs.~\ref{eq:11} we can also infer the effect of zealotry on the system: the effective noise increases with a term proportional to $N_1z_1/(N+z_1)$, while the herding decreases with the same factor.

\begin{figure}[!h]
 \centering
 \includegraphics[width=.23\textwidth]{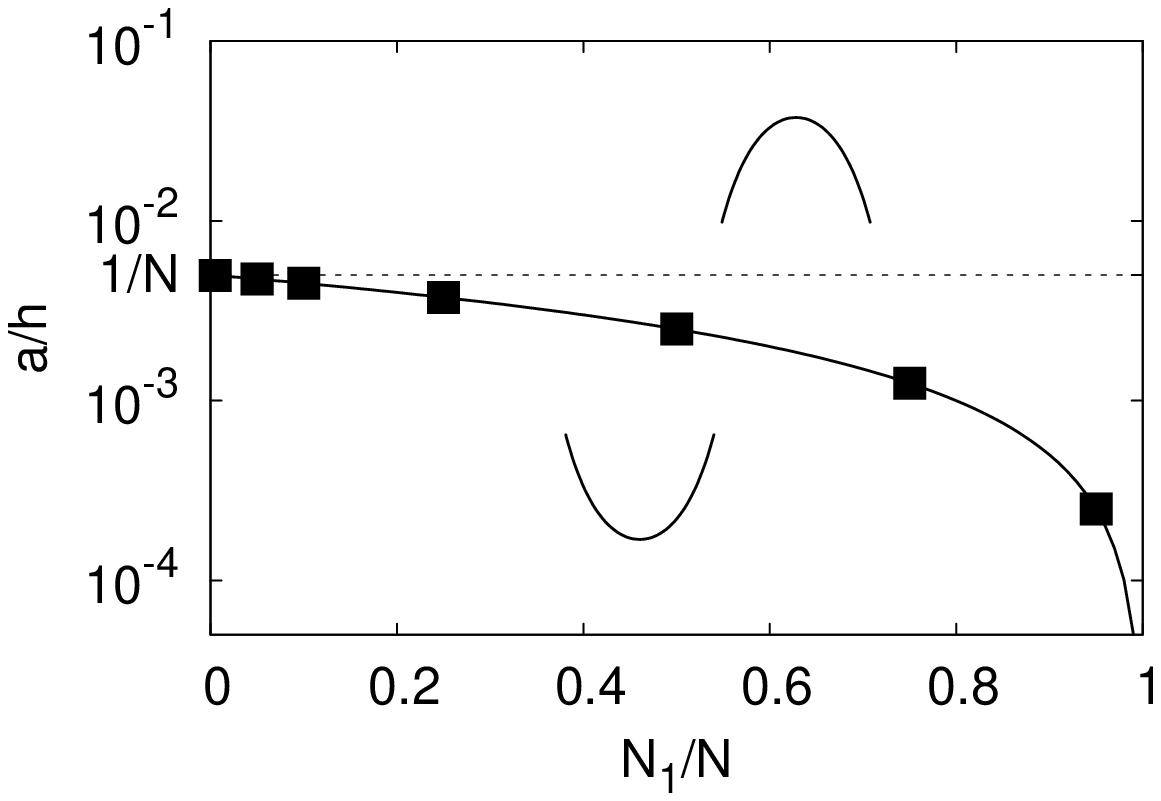}
 \includegraphics[width=.23\textwidth]{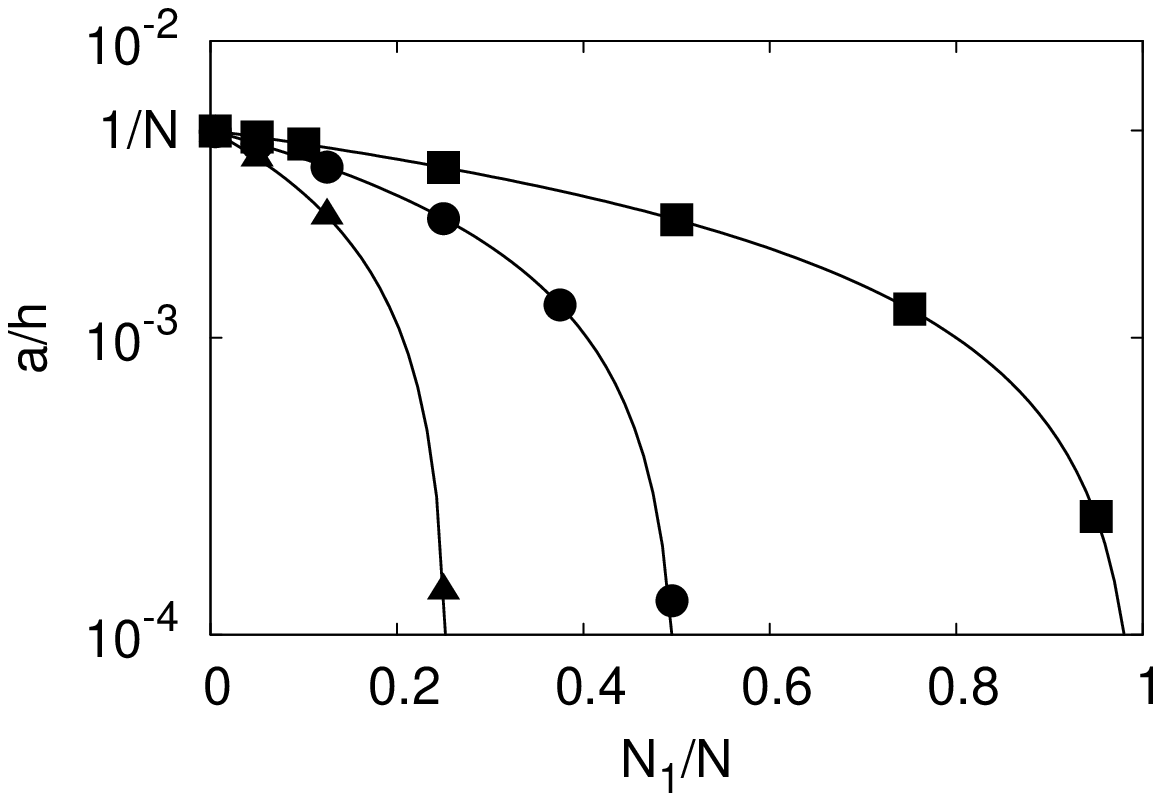}
 \caption{Phase diagrams with simulation (symbols) and theoretical (lines) results for a system with $N=200$ agents and $z_1^+=z_1^-=z_1/2$ with $z_1=0$ (dashed line), $2$ (squares), $4$ (circles), and $6$ (triangles).}
 \label{fig:9}
\end{figure}

The effective coefficients can be now used to derive the critical expression $(a/h)_c$, given by condition $a_b/h_b=1/N$, since the system at this approximation has the same phenomenology as that of the noisy voter model:
\begin{equation}
 \label{eq:12}
 (a/h)_c=\frac{1}{N}\left[1-\frac{(N+2)z_1}{2N(N+z_1)}N_1\right]
\end{equation}
(see Sec.~\ref{sec:5} for an alternative derivation based on a master--equation study).
The domain of validity of the different phases are better visualized by considering a phase diagram in the $(a/h,N_1/N)$ plane. In this diagram, Eq.~\eqref{eq:12} gives a critical line dividing the space parameters into two disjoint, symmetric bimodal and symmetric unimodal, regions, as shown in Fig.~\ref{fig:9}. Without the influence of zealots, i.e. with $z_1=0$, Eq.~\eqref{eq:12} is a horizontal line in the phase diagram, the dashed line of the left plot of Fig.~\ref{fig:9}. Upon increasing the number of zealots, the critical line moves toward the bottom of the diagram, making the SB phase narrow. In other words, for a given $N_1$ and by increasing $a/h$, the system may transient from the SB to the SU phases, at a smaller $a/h$ for larger $z_1$. Eventually, the number of zealots is so high so that if $N_1\ge N_1^*(z_1)$ the only feasible phase to the system is the SU phase, that is to say there is no value of $a/h$ for which the system can stay at the SB phase. The critical value $N_1^*(z_1)$ is given by imposing $(a/h)_c=0$,
\begin{equation}
 \label{eq:13}
 N_1^*(z_1)=\frac{2(N+z_1)}{(N+2)z_1}N,
\end{equation}
which is always larger than $N_1^{**}\equiv 2N/(N+2)\simeq 2$ and smaller than $N$ for $z_1> 2$. Hence, for $N_1<N_1^{**}$ the two phases are possible, regardless the number of zealots $z_1$, while for $N_1>N_1^{**}$ the SB phase disappears for $N_1>N_1^*(z_1)$.

\begin{figure}[!h]
 \centering
 \includegraphics[width=.3\textwidth]{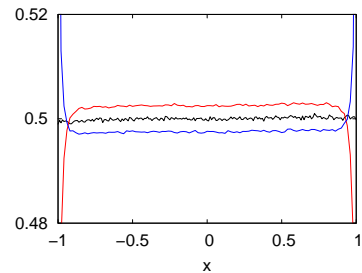}
 \caption{Probability distributions of the global magnetization (black) and partial magnetization of first community (blue) and second community (red) for a system with $N=200$, $N_1=N/2$, $z_1^+=z_1^-=1$, and $a/h=(a/h)_c$. While the global system is in a uniform phase (as indicated by the flatness of the pdf), the first community is in the symmetric bimodal phase, where the second community is in the symmetric unimodal phase.}
\label{fig:10}
\end{figure}

So far, we have focused on global properties of the system, disregarding specific features of the two communities. In fact, the fundamental assumption in the derivation of Eq.~\eqref{eq:11}, and hence Eqs. \eqref{eq:12} and \eqref{eq:13}, is that of same statistical properties of the two communities. But this is not the case in general: if $N_1$ is small enough, for example, the fluctuations of the magnetization of the first community are expected to be larger than that of the second one, see Appendix \ref{app:3} for a quantitative comparison. Moreover, there are cases close to the critical points, where the global magnetization has a uniform probability function (uniform phase) while the communities are each in a different phase, see Fig.~\ref{fig:10}. Nevertheless, the statistical differences between communities turn out to be irrelevant for the determination of the global behavior of the system, as the excellent agreement between theory and simulations shown in Fig.~\ref{fig:9} reveals.

\subsection{Unbalanced case}
\begin{figure}[!h]
 \centering
 \includegraphics[width=.4\textwidth]{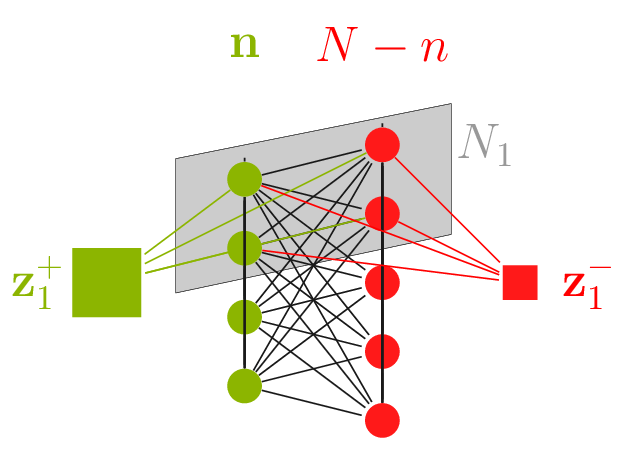}
 \caption{Schematic representation of the system of $M=2$ communities, only one directly influenced by different numbers of optimistic and pessimistic zealots.}
 \label{fig:11}
\end{figure}

For the unbalanced case, we consider a system with two communities $M=2$, with only one being influenced by optimistic and pessimistic zealots in different numbers $z_1^+\ne z_1^-$, as in Fig.~\ref{fig:11}.

\begin{figure}[!h]
 \centering
 \includegraphics[width=.23\textwidth]{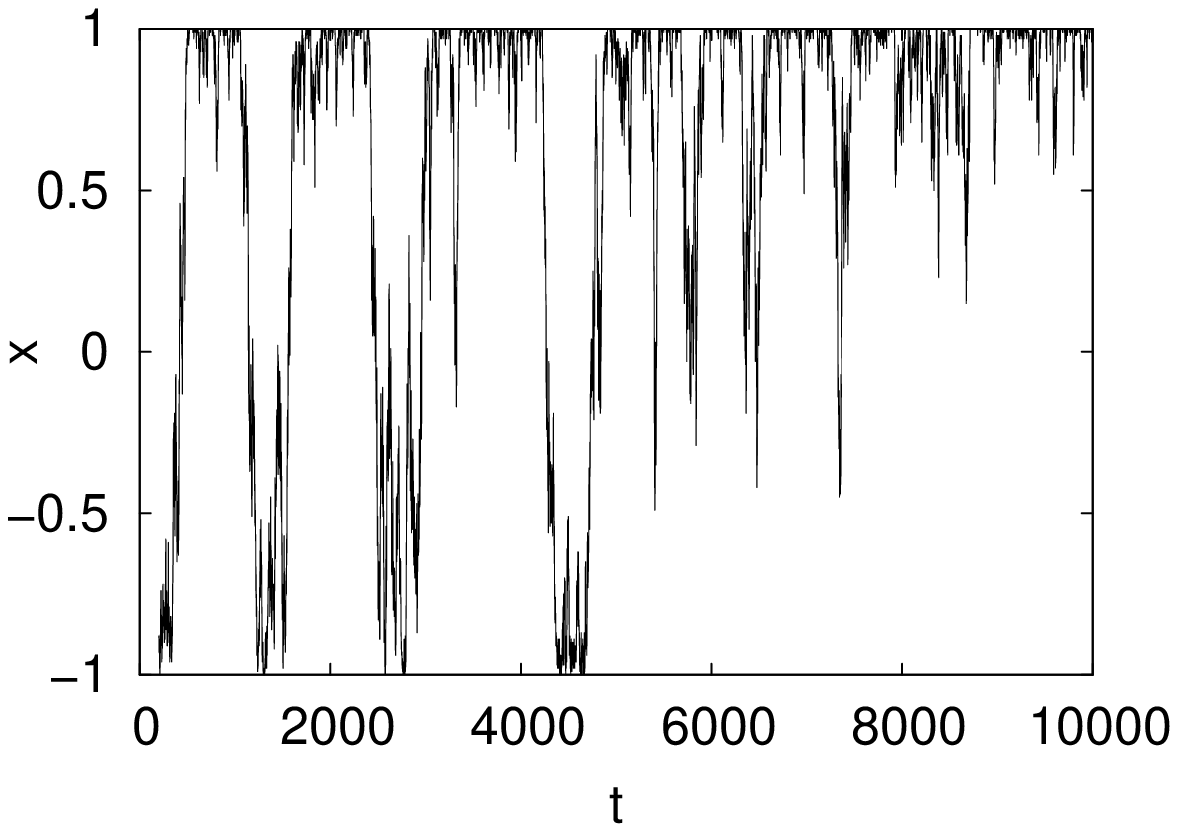}
 \includegraphics[width=.23\textwidth]{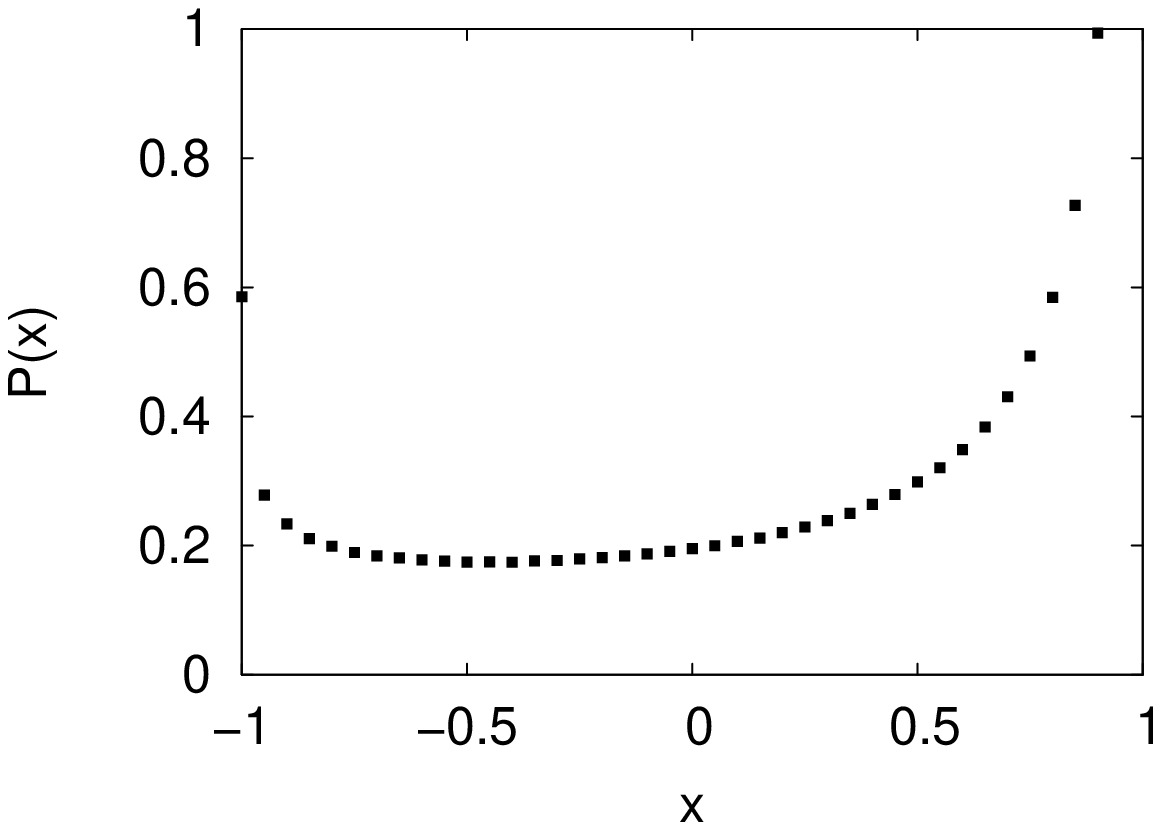}
 \includegraphics[width=.23\textwidth]{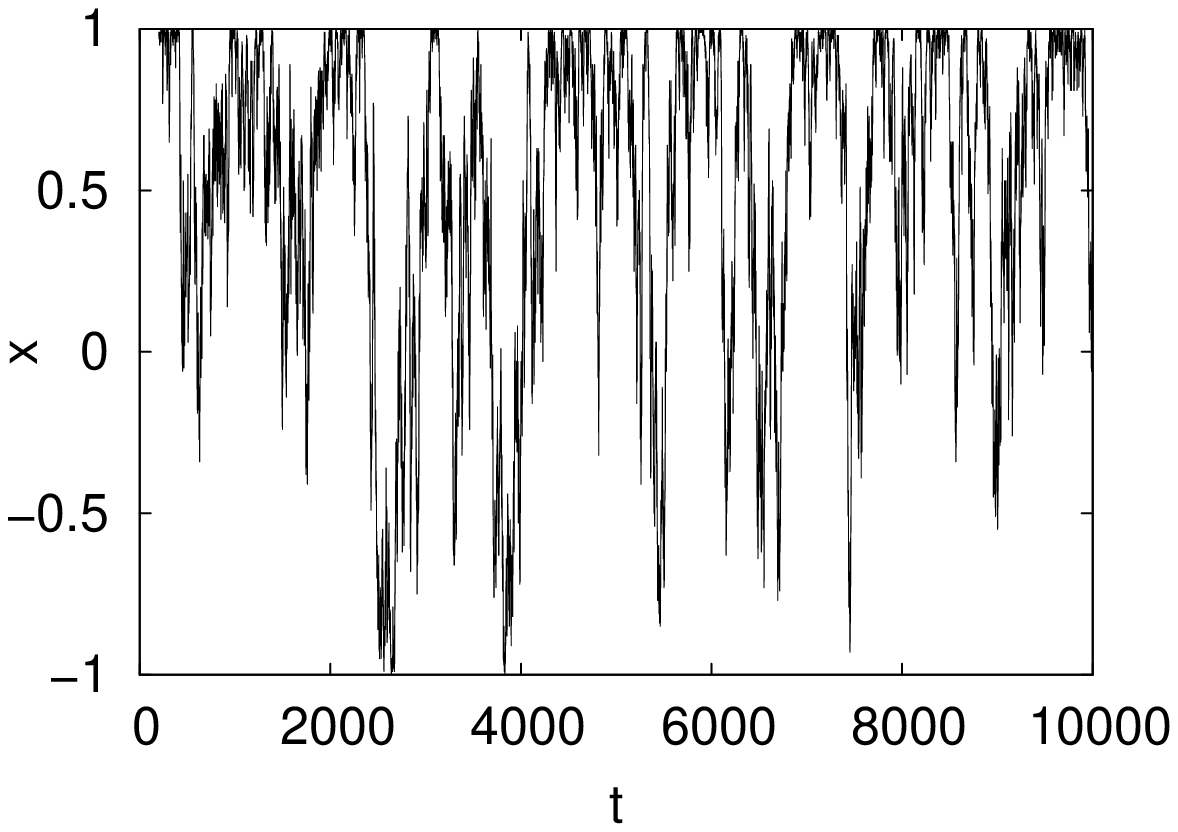}
 \includegraphics[width=.23\textwidth]{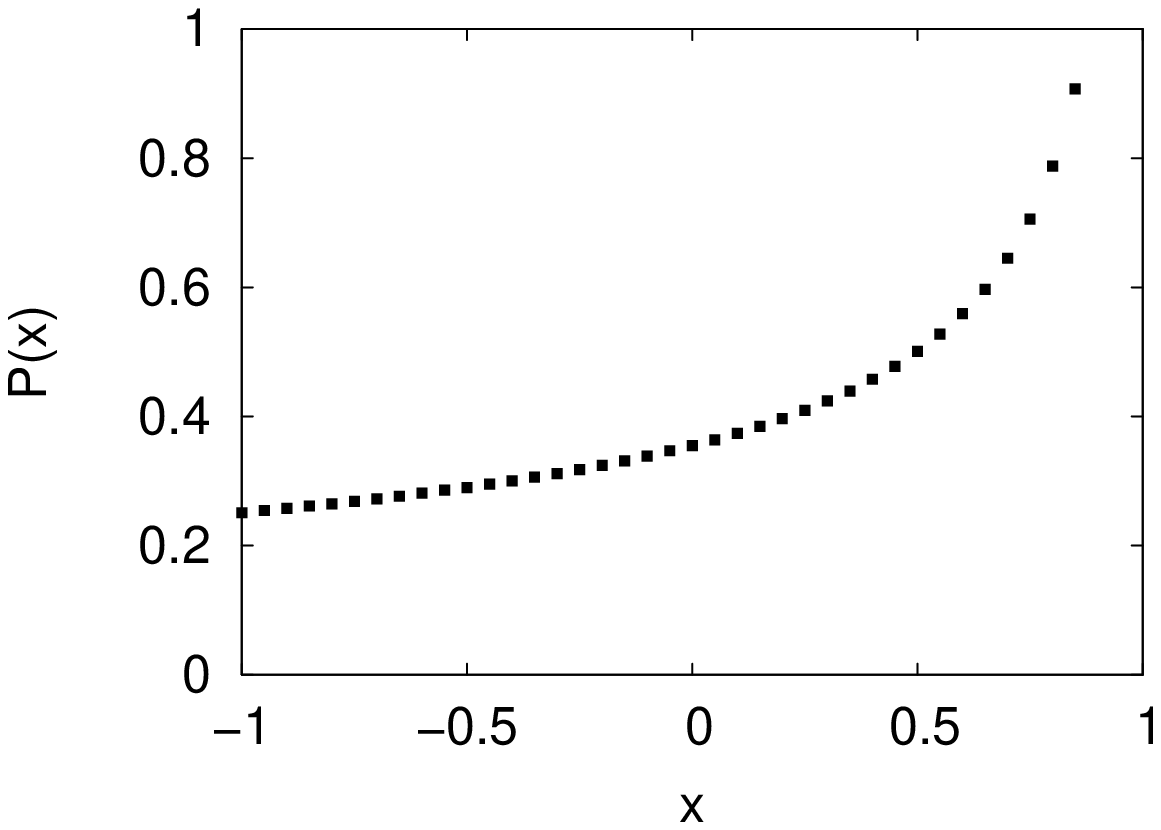}
 \includegraphics[width=.23\textwidth]{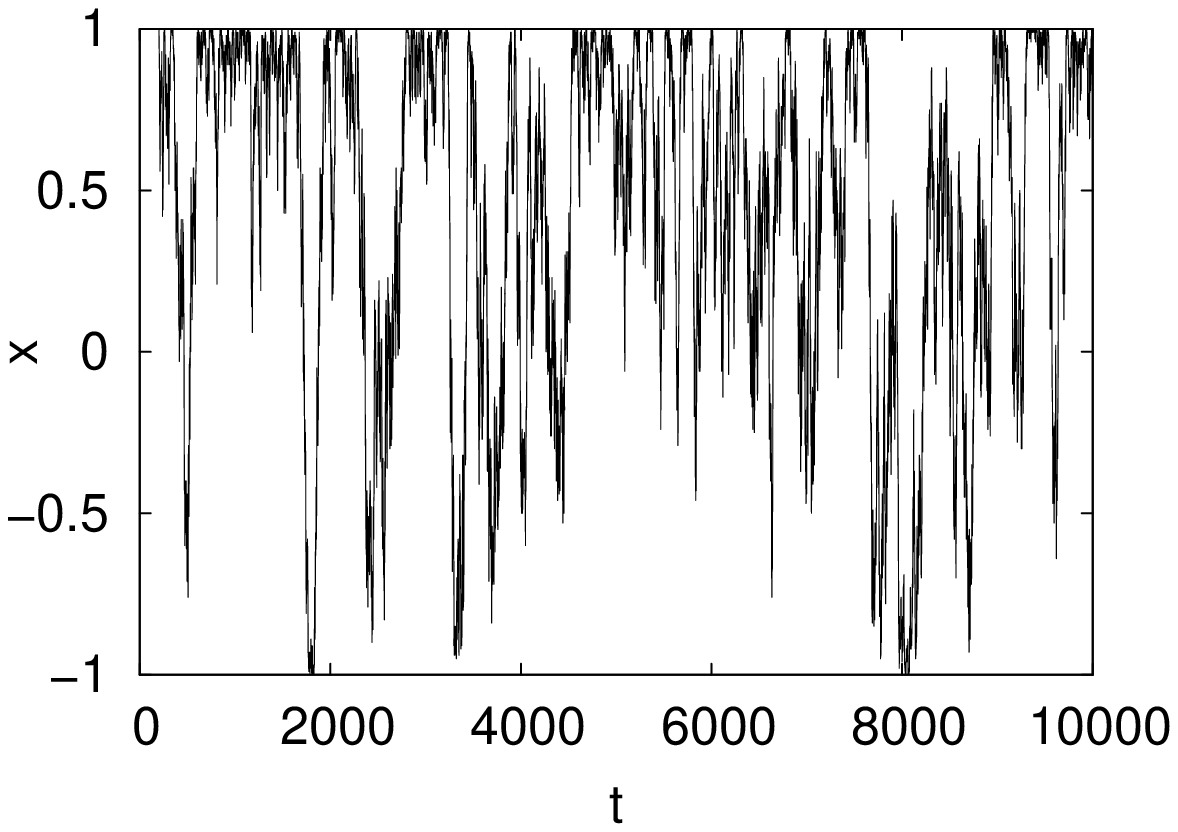}
 \includegraphics[width=.23\textwidth]{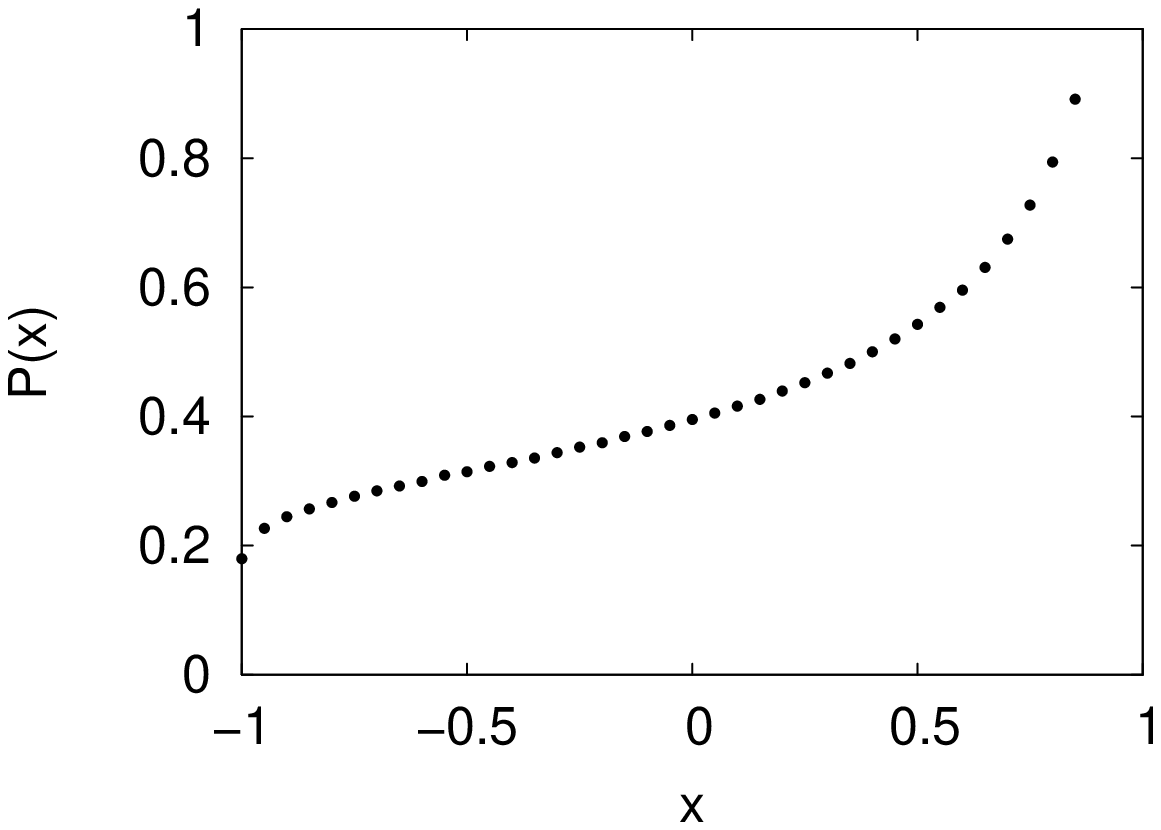}
 \includegraphics[width=.23\textwidth]{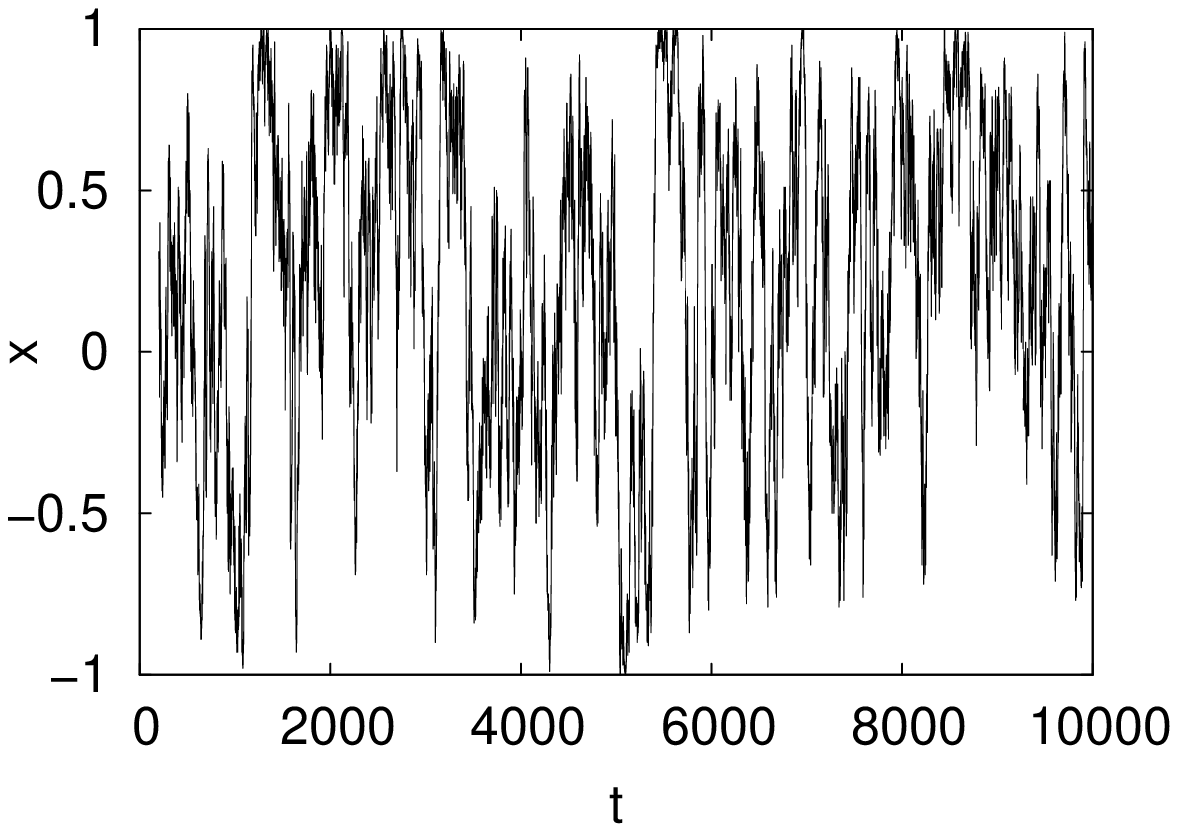}
 \includegraphics[width=.23\textwidth]{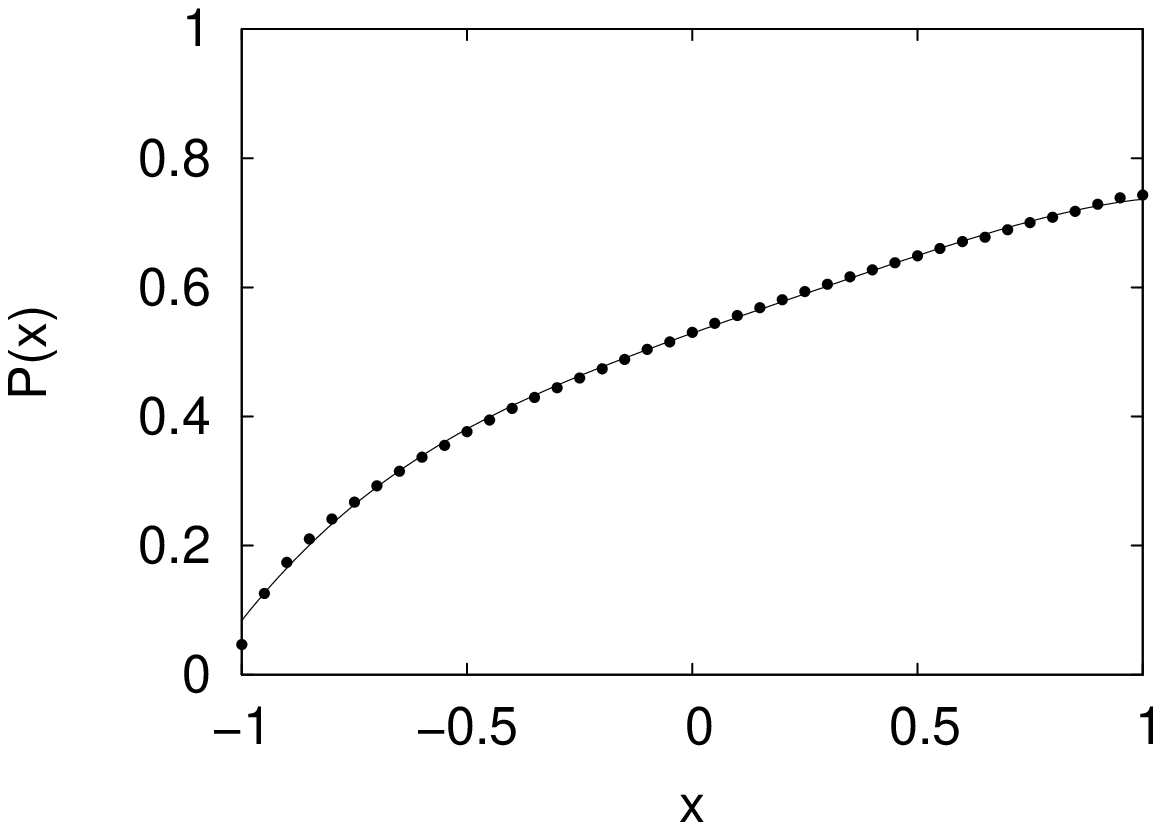}
 \includegraphics[width=.23\textwidth]{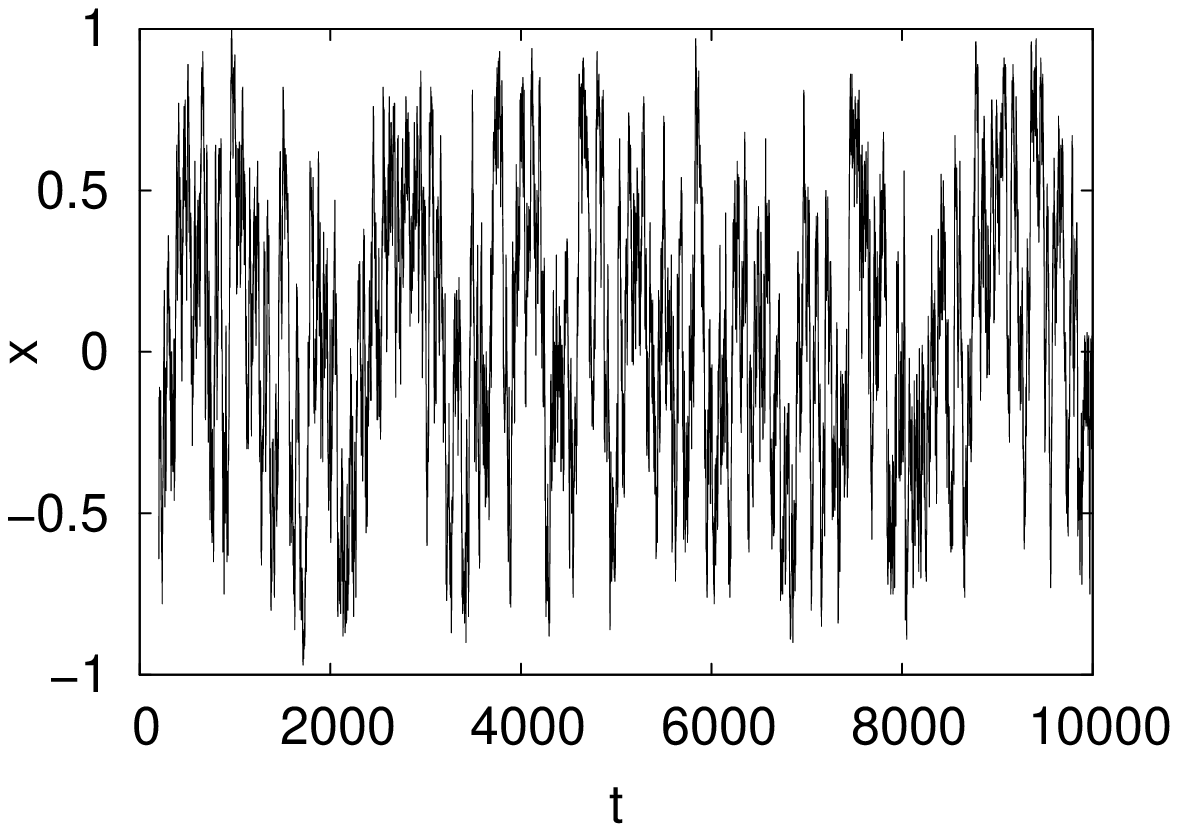}
 \includegraphics[width=.23\textwidth]{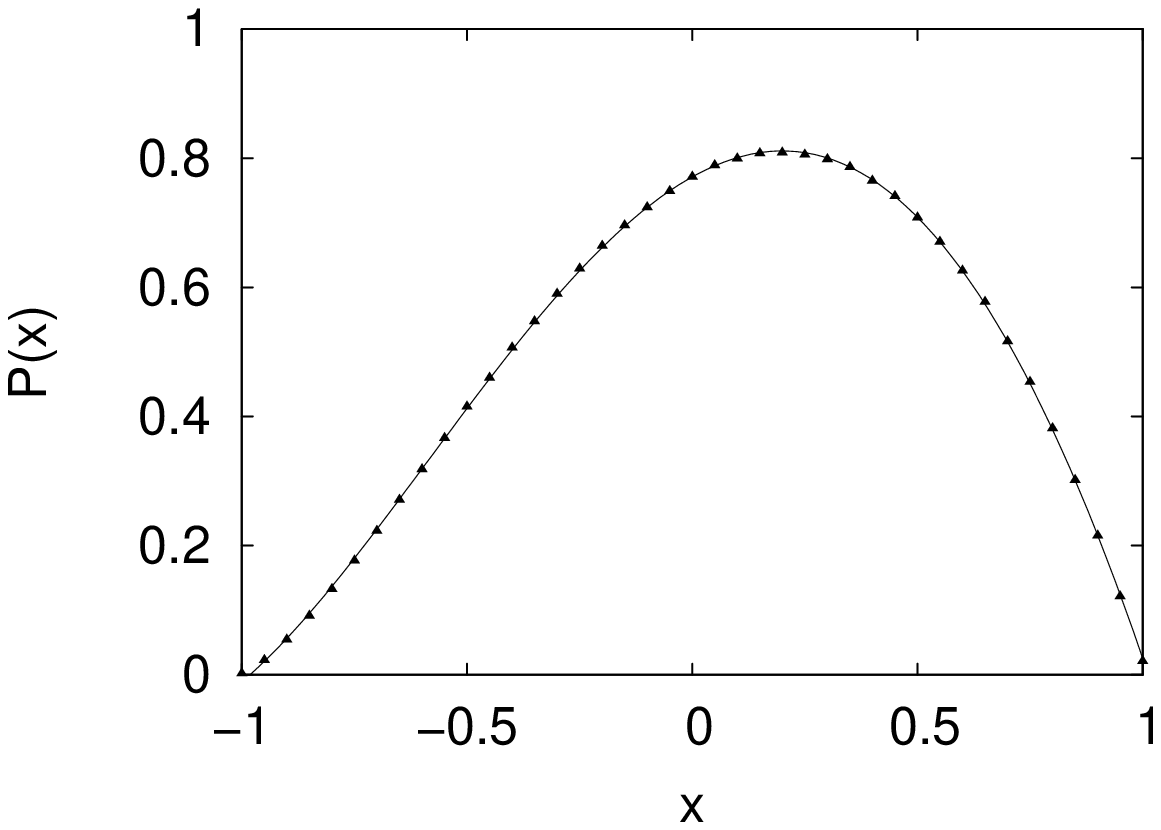}
 \caption{Simulations results for the trajectory $x(t)$ and probability function $P(x)$ of a system of $N=200$ agents, $N_1=N/2$, $z_1^+=1$, $z_1^-=0$, and $z_2=0$ for the following cases (from top to bottom in the figure) (i) $a/h< (a/h)_{c,1}$; (ii) $a/h=(a/h)_{c,1}$; (iii) $(a/h)_{c,1}<a/h< (a/h)_{c,2}$; (iv) $a/h=(a/h)_{c,2}$ ; and (v) $a/h> (a/h)_{c,2}$. Time is measured in units of $h^{-1}$. The solid lines on the last two plots of $P(x)$ are theoretical approximations derived in Appendix \ref{app:2}.}
 \label{fig:12}
\end{figure}

Following similar steps as in the balanced case, we provide first numerical simulations of the trajectories and their corresponding probability functions for the different phases the system may exhibit. As it is apparent from Fig.~\ref{fig:12}, the trajectories and probabilities are asymmetric, the opinion of the system tends to be that of the majority of zealots. Besides the extreme asymmetric and asymmetric unimodal phases already found for the one--community case, Fig.~\ref{fig:12} shows the existence of a new {\slshape asymmetric bimodal} (AB) phase characterized by the probability distribution having two relative maxima at the extreme values of the magnetization, see top row of Fig.~\ref{fig:12}. There are two critical values of $a/h$ that separates the three aforementioned phases: asymmetric bimodal for $(a/h)< (a/h)_{c,1}$, extreme asymmetric for $(a/h)_{c,1}<(a/h)< (a/h)_{c,2}$ and asymmetric unimodal for $(a/h)> (a/h)_{c,2}$. When reaching the transition point $(a/h)_{c,1}$ by increasing the value of $(a/h)$, the smallest relative maximum of $P(x)$ (located at the value of the magnetization opposite to the one preferred by the majority of zealots) becomes a relative minimum. Analogously, the absolute maximum at $x=\sign(\Delta z_1)$ becomes a relative minimum when reaching $a/h<(a/h)_{c,2}$ by increasing the value of $(a/h)$.

By a similar reasoning to the one used in Subsec.~\ref{sub:balanced} the effective coefficients are
\begin{equation}
\label{eq:14}
\begin{split}
 &a_u^\pm=a_b\pm\frac{N_1\Delta z_1}{2N(N+z_1)}h, \\
 &h_u=h_b,
\end{split}
\end{equation}
with $a_b$ and $h_u$ given by Eq.~\eqref{eq:11}. These rates are now used to compute the critical lines according to the general theory described in Appendix \ref{app:1} for a one-community system:
\begin{equation}
 \label{eq:15}
 \begin{split}
 (a/h)_{c,i}=&\frac{1}{N}\left[1-\frac{(N+2)z_1}{2N(N+z_1)}N_1\right]\\ &+(-1)^i\frac{(N+1)N_1}{2N(N-1)(N+z_1)}|\Delta z_1|,
 \end{split}
\end{equation}
for $i=1$ and $2$. Equation \eqref{eq:15} coincides with \eqref{eq:9} for $N_1=N$ and with Eq.~\eqref{eq:12} for $\Delta z_1=0$, as expected for consistency. This approach is also consistent with the more general one of Sec. \ref{sec:5}.

\begin{figure}[!h]
 \centering
 \includegraphics[width=.23\textwidth]{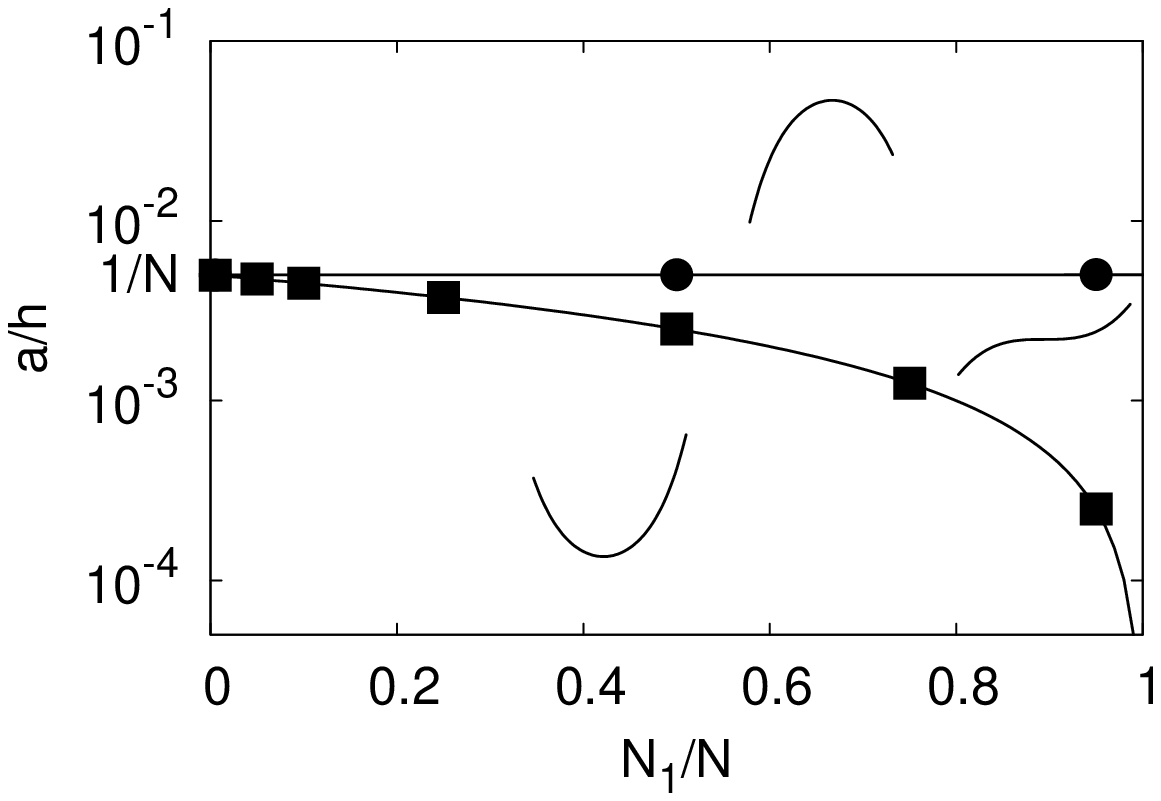}
 \includegraphics[width=.23\textwidth]{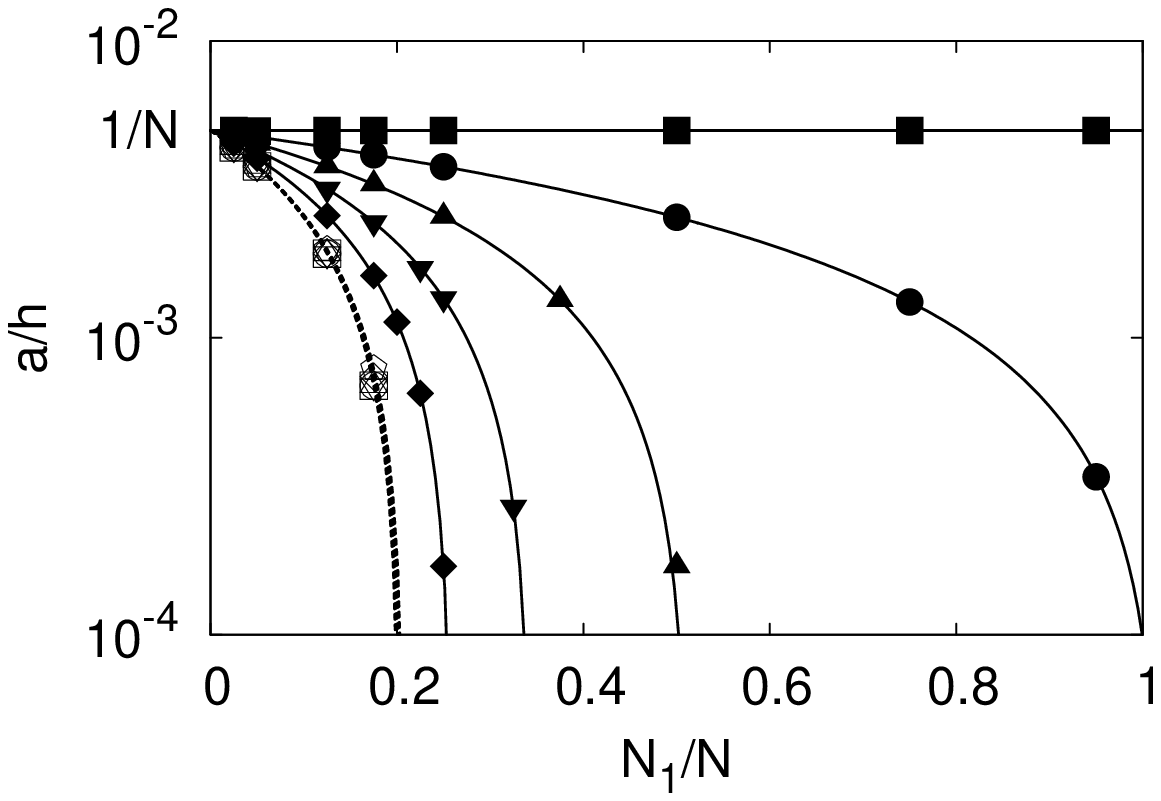}
 \caption{Phase diagrams for a system of $N=200$ agents. Left: $z_1^+=1,\ z_1^-=0$. Right: $z_1^+=5,\ z_1^-=0$ (squares), $1$ (up circles), $\dots,5$ (pentagon).}
 \label{fig:14}
\end{figure}

In Fig.~\ref{fig:14} we show the region of existence of the different phases in the parameter space $(a/h, N_1/N)$ for some representative situations. The phase space is divided by the two critical lines $(a/h)_{c,1}$ and $(a/h)_{c,2}$ into three disconnected regions, corresponding to the AB (bottom), EA (center), and AU (top) phases. In general, the positions of the two critical lines on the phase diagram depend very differently on the total number of agents $N$ and on the number of zealots $z_1^+$ and $z_1^-$. For the specific values of the right plot of Fig.~\ref{fig:14}, while the critical line $(a/h)_{c,1}$ does not change very much with $\Delta z_1$, as far as it is small compared to $N$, the critical line $(a/h)_{c.2}$ moves from almost a horizontal line for $z_1=|\Delta z_1|$ to $(a/h)_{c,1}$ for $\Delta z_1=0$. When the number of optimistic and pessimistic zealots becomes equal, the two critical lines coincide and we recover the results of the balanced case.

\begin{figure}[!h]
 \centering
 \includegraphics[width=.23\textwidth]{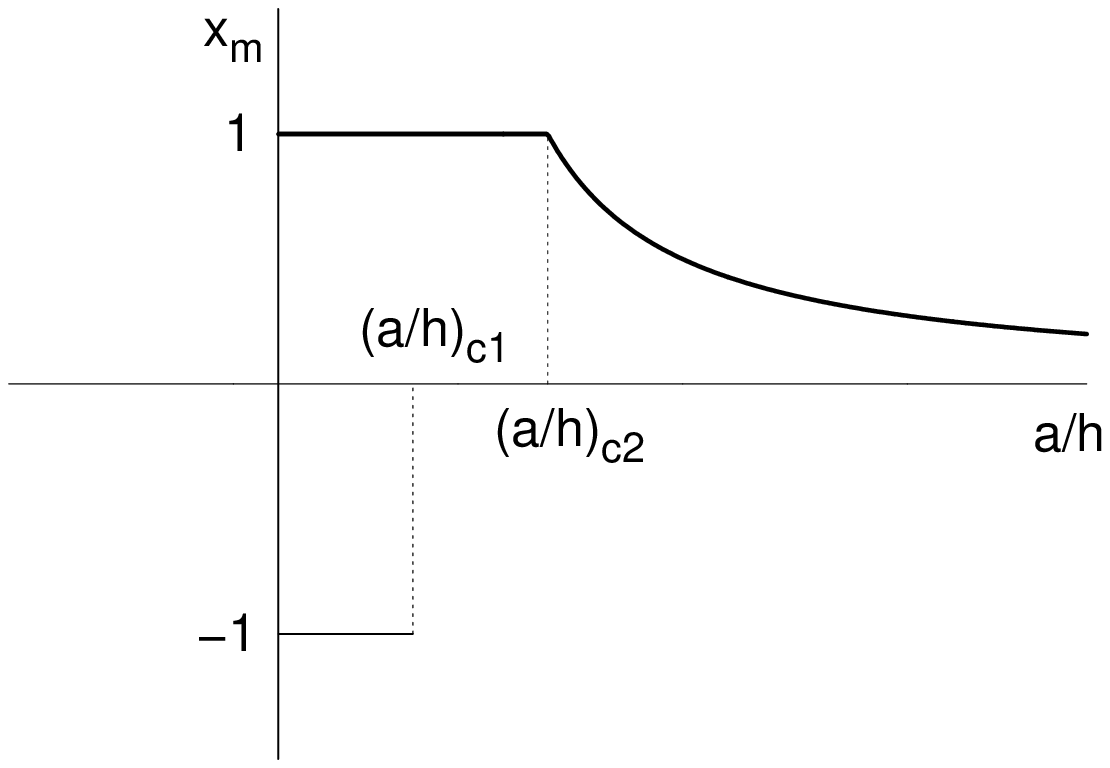}
 \includegraphics[width=.23\textwidth]{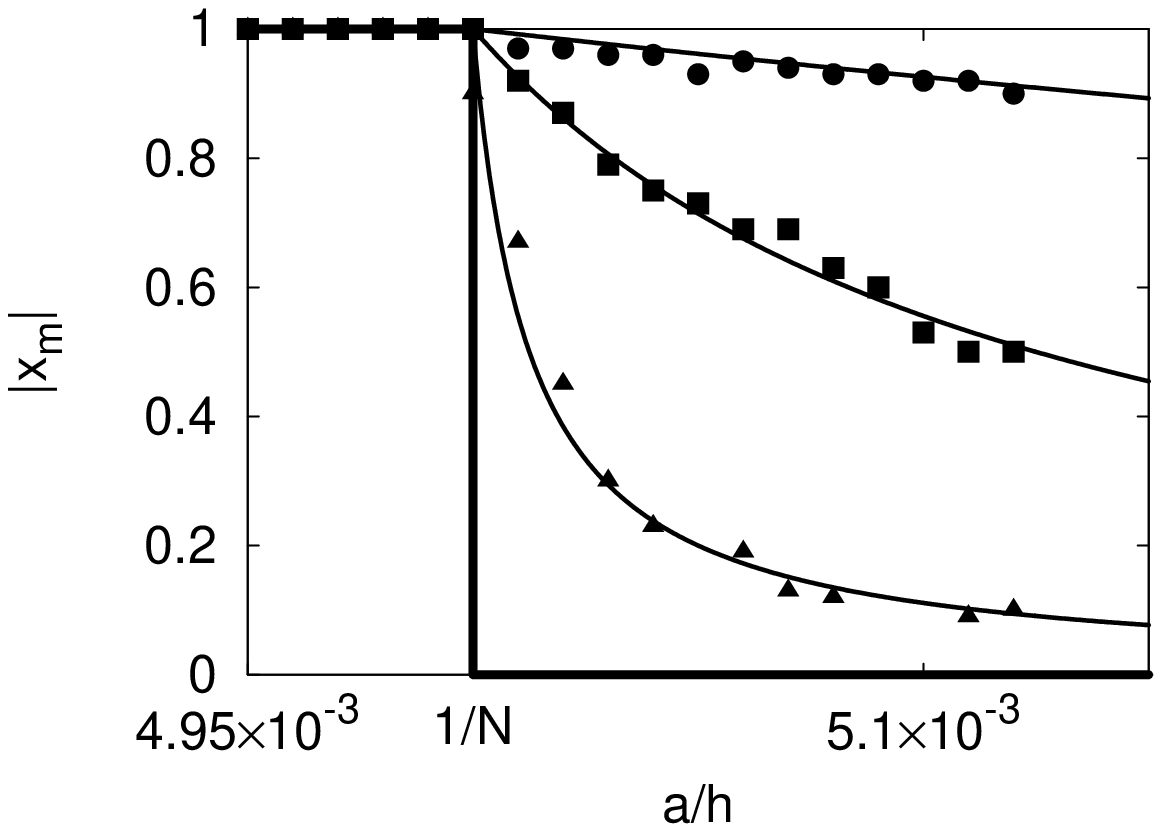}
 \caption{Left: schematic representation of the maxima of the probability function for a case of $z_1^+>z_1^-$. Right: numerical (symbols) and theory (lines) results for the absolute maximum for a system of $N=200$ agents, $z_1^+=1$, $z_1^-=0$, and $N_1=N/200, N/20, 2, 0$ (from top to bottom).}
 \label{fig:13}
\end{figure}

As for the case of one community, the absolute maximum of the probability function changes continuously with $a/h$, even at the critical point $(a/h)_{c,2}$. In contrast, the relative maximum disappears discontinuously at $(a/h)_{c,1}$, see Fig.~\ref{fig:13}.

\section{General case \label{sec:5}}
After having discussed particular cases in the previous sections, we focus now on the phase diagram and its critical lines for a general case of $M$ communities with different numbers of zealots affecting different communities. After a systematic numerical study, we confirm the general unbalanced case of the previous scenarios, namely that the system may stay in the asymmetric bimodal, extreme asymmetric or asymmetric unimodal phases, that correspond to three regions separated by the critical lines $(a/h)_{c,1}$ and $(a/h)_{c,2}$, in a way analogous to the unbalanced case of $M=2$. This is the most general situation, since it contains the balanced ones as a limit: the EA phase disappears, since $(a/h)_{c,1}$ and $(a/h)_{c,2}$ become the same line, and AB and AU phases become SB and SU phases, respectively. Thus, in order to unveil the structure of the phase diagram, the only thing needed is to determine the location of the two critical lines for which we apply here a line of reasoning based on an analysis of the master equation.

The critical lines of the phase diagram correspond to values of the parameters of the system where the probability $P(n)$ exhibits some peculiarities for $n=0,N$. In general, however, $P(n)$ does not obey an autonomous equation, and we have to consider first the probability of finding the system at a state $S=\{n_1,\dots,n_M\}$, the function $p(S)=p(n_1, \dots, n_M)$. It obeys the following master equation valid for steady--state conditions
\begin{equation}
 \label{eq:16}
 \begin{split}
 \sum_{k=1}^M & \left[ (E^+_k-1)\pi_k^-(S) p(S)\right. \left.+(E_k^--1)\pi_k^+(S)p(S)\right]=0,
 \end{split}
\end{equation}
where we have made explicit the dependence of the rates Eq.~\eqref{eq:1} on the state of the system, and $E_i^\pm$ are operators acting on an arbitrary state function $f(S)$ as $E_i^\pm f(S)=f(n_1,\dots,n_i\pm 1,\dots,n_M)$.

For $n=0$, the only possible state is $S=(0,\dots,0)$ for which the master equation implies
\begin{equation}
 \label{eq:17}
 \begin{split}
 &\sum_{k=1}^M \pi_k^-(\{0,\dots, \underbrace{1}_{k}, \dots, 0\}) p(\{0,\dots, \underbrace{1}_{k}, \dots, 0\}) \\
 &=\left(\sum_{k=1}^{M}\pi_k^+(\{0,\dots,0\})\right) p(\{0,\dots,0\}).
 \end{split}
 \end{equation}
For the determination of the critical lines we now make the approximation that
\begin{equation}
 \label{eq:18}
 p(\{0,\dots, \underbrace{1}_{k}, \dots, 0\})\simeq \frac{N_k}{N} P(n=1),
\end{equation}
which assumes that if there is only one optimistic agent, the probability that he belongs to community $k$ is proportional to the population of that community (an assumption trivially satisfied for $M=1$). This way, since $p(\{0,\dots,0\})=P(n=0)$, we have
\begin{equation}
 \label{eq:19}
 P(n=1)\simeq N\frac{\sum_{k=1}^{M}\pi_k^+(\{0,\dots,0\})}{\sum_{k=1}^M N_k \pi_k^-(\{0,\dots, \underbrace{1}_{k}, \dots, 0\})} P(n=0).
\end{equation}
The critical line $(a/h)_{c,1}$ appears when $P(n=0)$ changes from local maximum to local minimum, or $P(n=1)=P(n=0)$. After replacing the rates given by Eq.~(\ref{eq:1}) in Eq.~(\ref{eq:19}) we obtain
\begin{equation}
 \label{eq:20}
 (a/h)_{c,1}=\frac{1}{N(N-1)}\sum_{k=1}^M \frac{N_k}{N+z_k}\left[z_k^--1-N(z_k^+-1)\right].
\end{equation}
By symmetry considerations, the other critical line is
\begin{equation}
 \label{eq:21}
 (a/h)_{c,2}=\frac{1}{N(N-1)}\sum_{k=1}^M \frac{N_k}{N+z_k}\left[z_k^+-1-N(z_k^--1)\right].
\end{equation}
As expected, by taking $z_k^+=z_k^-$, the critical lines coincide.

\section{Discussion and conclusions \label{sec:6}}

In this work, we have analyzed the influence of agents that never change their state (zealots) on the global properties of the noisy voter model. Only simple situations of all--to--all connection among voters have been considered, but still allowing different communities of agents to be directly influenced by a different number of zealots. In the zealots-free case, it is known that when increasing the noise to herding ratio the noisy voter model displays a finite-size transition from a symmetric bimodal phase (where consensus is the norm) to a symmetric unimodal phase with coexistence of opinions. As explained in Sec. \ref{sec:2}, the dynamics of the voters with zealotry is equivalent to that of heterogeneous (different noise and herding constants) noisy voters without zealots. This equivalence between models allows a straightforward explanation of how zealotry acts on the system for some simple cases.

In the case of a balanced number of optimistic and pessimistic zealots with global influence (where zealots act upon all agents), it turns out that the consensus (symmetric bimodal, SB)  phase disappears and the system is always in the symmetric unimodal (SU) phase. This result shows the dramatic influence of the zealots, even for their lowest possible number ($z=2$). If the balanced number of zealots influence only a fraction $N_1/N$ of the total population, then the symmetric bimodal phase can still be present as far as the fraction $N_1/N$ is smaller than some critical value as given by Eq.~\eqref{eq:13} and summarized in Fig.~\ref{fig:9} for a particular case. { Similar results have been found on a kinetic model of opinion dynamics \cite{crblan14}}.

In the unbalanced scenario where unequal numbers of optimistic and pessimistic zealots influence the whole population, the main result is the appearance of asymmetric phases which tilt the distribution of opinions towards the one favored by the majority of zealots. Increasing the noise to herding ratio the model displays a transition from an extreme asymmetric (EA) phase (where the maximum of the probability distribution occurs at the consensus value favored by the zealots) to an asymmetric unimodal (AU) phase where the maximum, being still tilted towards the zealot-favored opinion, is located far from the extreme consensus opinion, indicating again a strong qualitative influence of the zealots on the system. The extreme asymmetric phase does not exist for a sufficiently large population of zealots, i.e. $z>2+|\Delta z|$, being  $\Delta z$ the difference between the number of zealots of each type. If the unbalanced number of zealots acts only upon a sufficiently small fraction of the population $N_1/N$, then a new, asymmetric bimodal (AB), phase can appear. This phase is characterized by a probability distribution showing relative maxima at both consensus states.

In the general case of several communities, $k=1,\dots,M$, and sets of zealots acting on a particular community, the phenomenology is similar to the one described above.  If  all communities suffer the influence of a balanced number of zealots $z_k^+=z_k^-,\,\forall k$, then the possible outcomes are the symmetric bimodal (SB) or unimodal (SU) phases, depending on the particular value of the noise to herding ratio, as determined by Eq.~\eqref{eq:20}. This is also the situation when the optimistic--pessimistic balance is conserved, for instance if $N_1=N_2=N/2$, and $z_1^+=z_2^-$ and $z_1^-=z_2^+$. In other cases, three asymmetric phases (extreme EA, unimodal UA or bimodal BA) can be present in a region of the parameter space as defined by the two lines $(a/h)_{c,1}$ and $(a/h)_{c,2}$ given by Eqs.~(\ref{eq:20},\ref{eq:21}), as shown in Fig.~\ref{fig:14} in a particular case. Again, not all the phases are compatible with all possible numbers of zealots and of their links.

It is also interesting to analyze the results of the present work in terms of the competition between zealots of different opinions willing to have the largest possible number of agents in the same state as the zealot. It is clear that the best situation for the zealot is to break the symmetry of the problem going to the EA phase, but if the symmetry cannot be broken, the best strategy is rather counterintuitive: In this case the best situation for each zealot is the bimodal phase in which most of the population coincides with the state of one zealot for long periods of time. To achieve this result, the strategy of the zealot is to interact with a small number of agents ($N_1<N_1^*$, Eq.~\eqref{eq:13}) to make sure that for a low enough value of the ratio of the noise to herding parameters, the system remains in the bimodal phase (Fig.~\ref{fig:9}). The reason for that can be understood in our mapping of zealots into a heterogeneous noisy voter model: the action of zealots plays the role of an effective noise that tends to bring the system into the unimodal phase.

{In conclusion, we have shown that, in general, upon introducing zealots in an homogeneous population of (noisy) voters, the dynamics of the system changes drastically: a breaking of symmetry can be induced, and new phases describing the global behavior of the system may appear. Although the aforementioned results have been obtained at the mean--field level, where all--to--all links connect all voters, we expect a similar phenomenology for more realistic, high dimensional networks, which are the most representative in social systems. A detailed analysis on the impact of the network dimension and topology remains an open question.}

\acknowledgments
We acknowledge financial support from Ministerio de Economía y Competitividad (MINECO) and Fondo Europeo de Desarrollo Regional (FEDER) under project ESOTECOS FIS2015-63628-C2-1-R.

\appendix

\section{Closed system of equations for the moments \label{app:3}}
Consider the master equation for the probability function $P(S)$ with the general rates in \eqref{eq:1} or \eqref{eq:2} written as
\begin{equation}
  \label{eq:3a1}
  \frac{d P(S)}{d t}=J[P],
\end{equation}
where $J[P]$ stands for the l.h.s of Eq.~\eqref{eq:16}. The mean values of a generic quantity $n_1^{c_1}n_2^{c_2}\dots n_M^{c_M}$, for given integer values $c_1,\dots,c_M$, can be obtained by multiplying Eq.~\eqref{eq:3a1} by the same quantity and summing over all possible states:
\begin{equation}
  \label{eq:3a2}
  \frac{d}{dt}\medio{n_1^{c_1}n_2^{c_2}\dots n_M^{c_M}}=\sum_{\medio{S}}n_1^{c_1}n_2^{c_2}\dots n_M^{c_M}J[P,S].
\end{equation}
Taking into account the fact that $P$ vanishes for nonphysical values of $S$ and the explicit form of the rates, it is not difficult to have the following general property
\begin{equation}
  \label{eq:3a3}
  \begin{split}
    & \sum_{\medio{S}}f(n_1,\dots,n_M)E^\pm_k[\pi_k^\mp(S)P(S)] \\
    & =  \sum_{\medio{S}}\pi_k^\mp(S)P(S)E^\mp_kf(n_1,\dots,n_M)
  \end{split}
\end{equation}
for any function $f$. The latter expression allows us to write Eq.~\eqref{eq:3a2} as
\begin{equation}
  \label{eq:3a4}
  \begin{split}
    &\frac{d}{dt}\medio{n_1^{c_1}n_2^{c_2}\dots n_M^{c_M}} = \\
    &\sum_{k=1}^M\medio{\left[\pi_k^+(E^+_k-1)+\pi_k^-(E^-_k-1)\right]n_1^{c_1}n_2^{c_2}\dots n_M^{c_M}}.
  \end{split}
\end{equation}
It turns out that the r.h.s. of Eq.~\eqref{eq:3a4} is a polynomial of degree $c_1+\dots+c_M$. This is because
\begin{equation}
  \label{eq:3a5}
  \begin{split}
    &\left[\pi_k^+(E^+_k-1)+\pi_k^-(E^-_k-1)\right]n_k^{c_k} \\ &=c_k(\pi_k^+-\pi_k^-)n_k^{c_k-1}+\mathcal{O}(n_k^{c_k-1})
  \end{split}
\end{equation}
is of degree $c_k$, since $\pi_k^+-\pi_k^-$ is of degree one, as it is apparent from Eq.~\eqref{eq:1}.

\begin{figure}[!h]
 \centering
 \includegraphics[width=.23\textwidth]{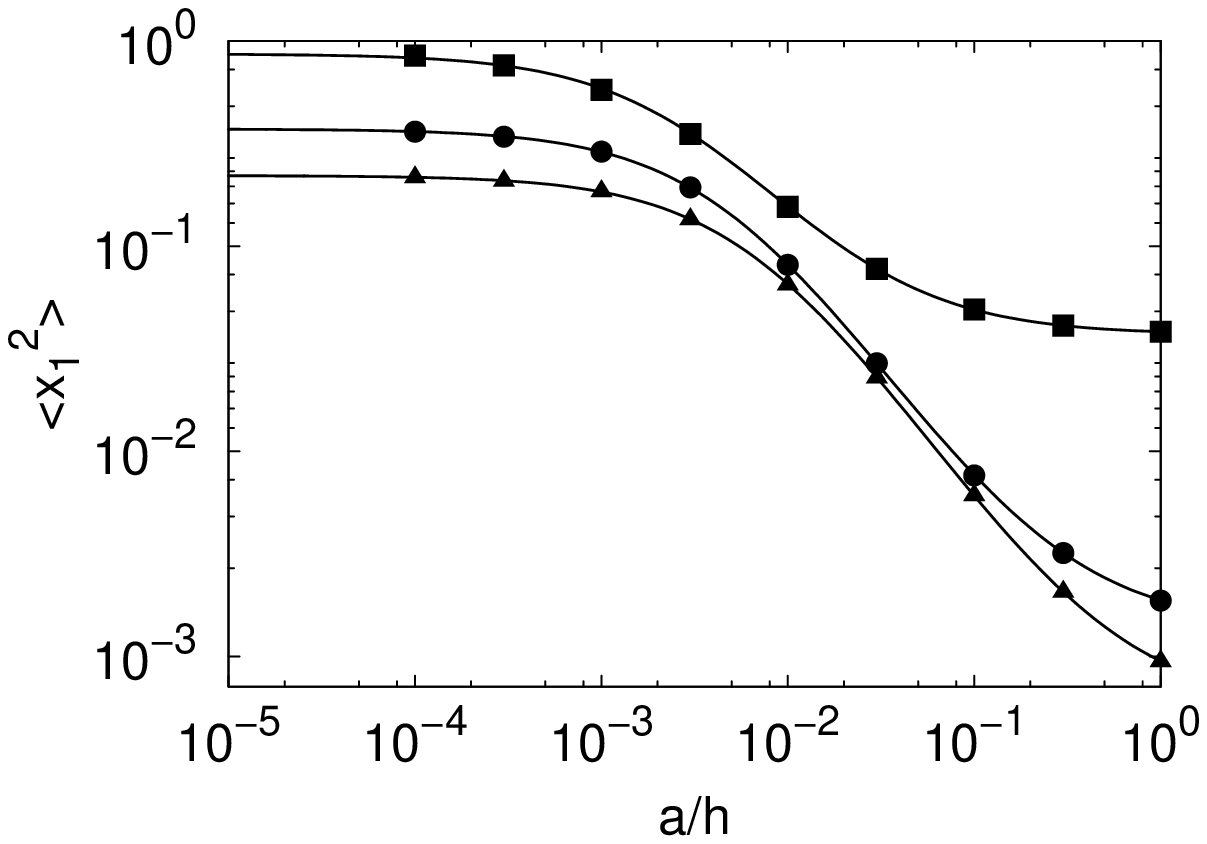}
 \includegraphics[width=.23\textwidth]{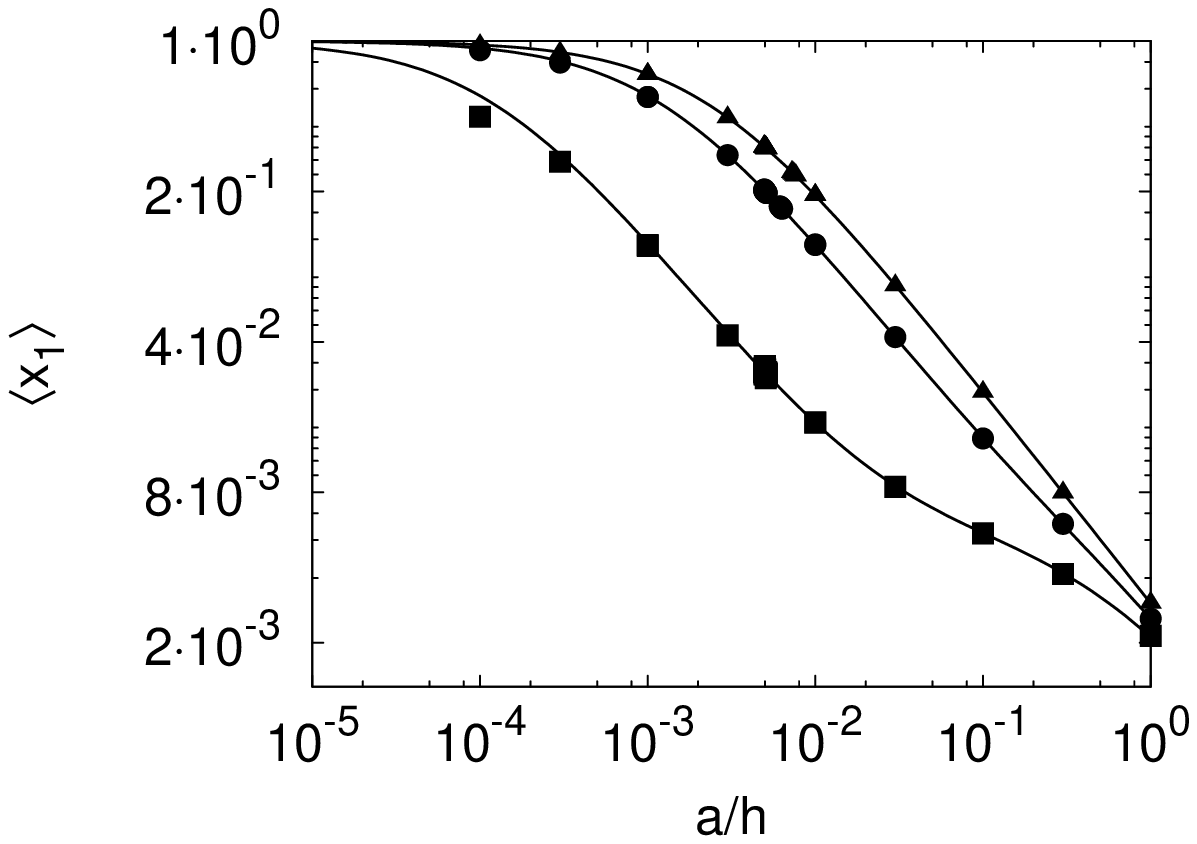} \\
 \includegraphics[width=.23\textwidth]{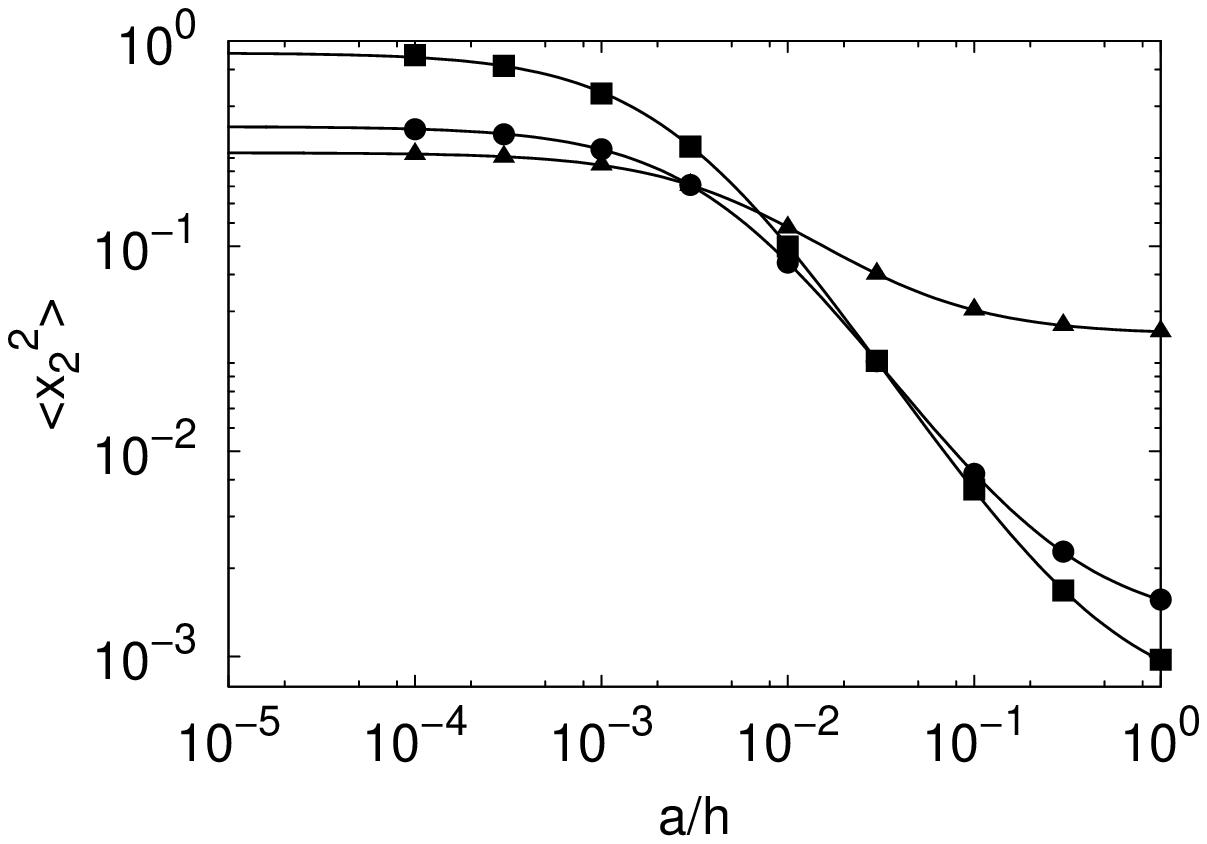}
 \includegraphics[width=.23\textwidth]{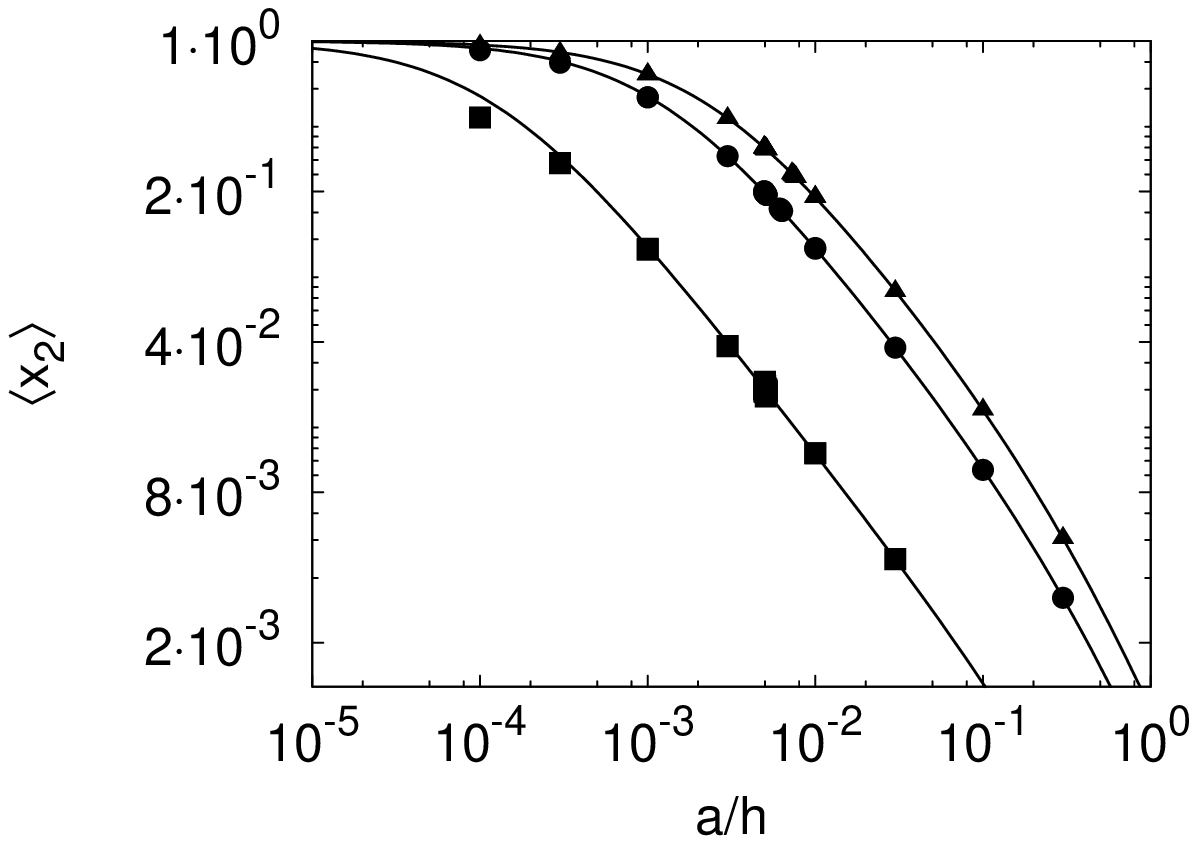} \\
 \includegraphics[width=.23\textwidth]{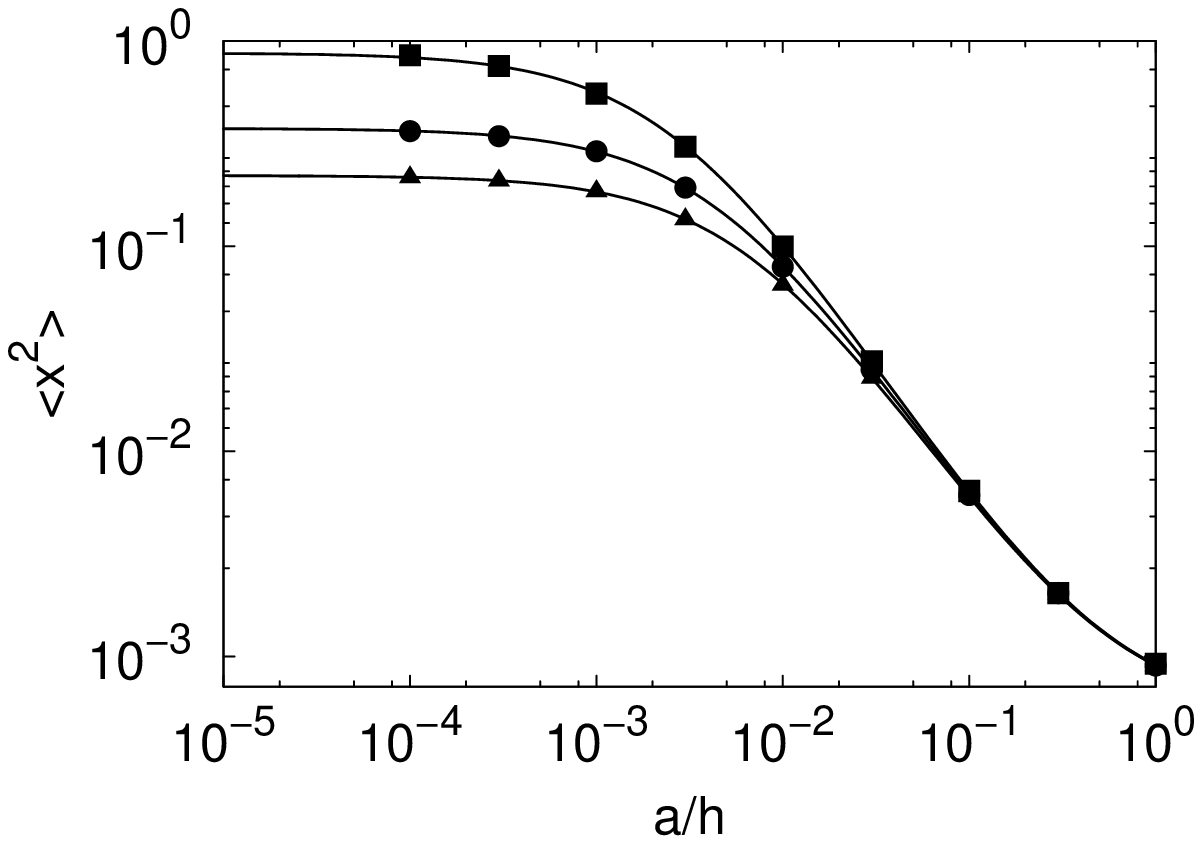}
 \includegraphics[width=.23\textwidth]{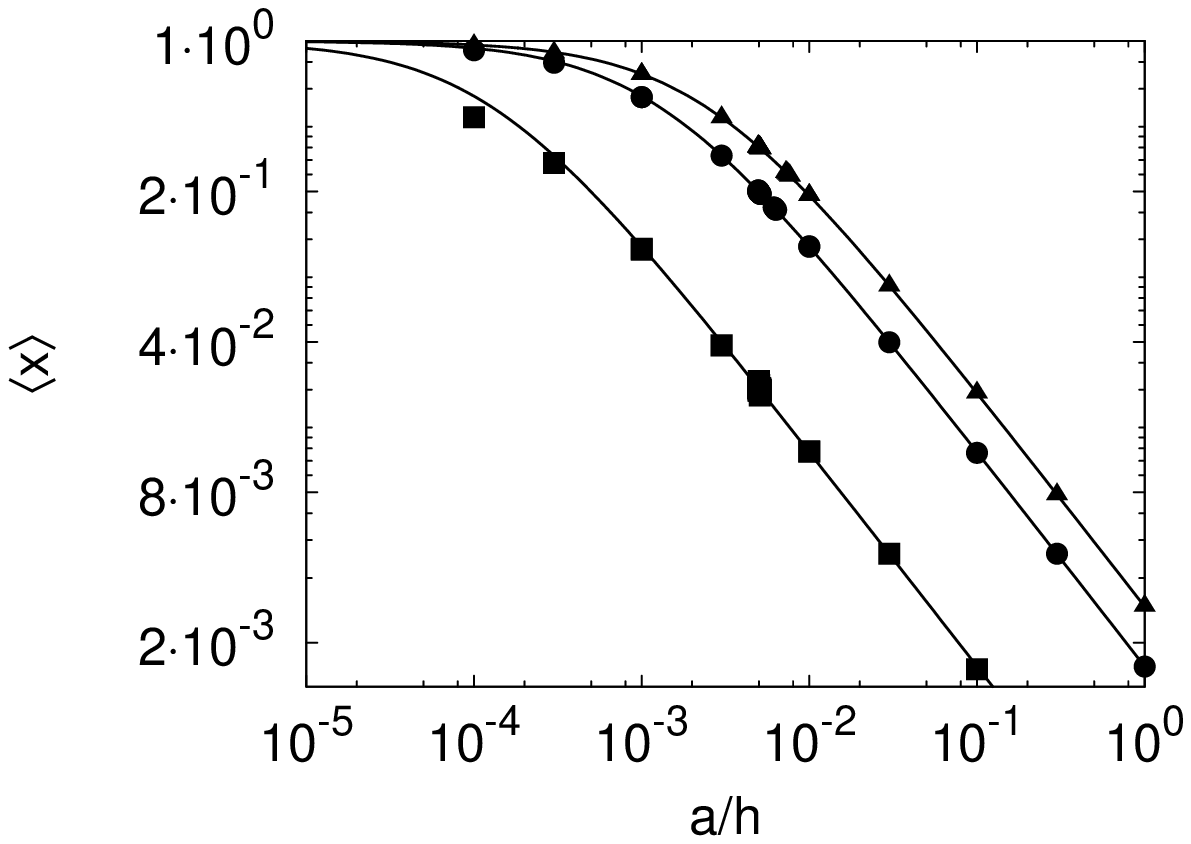} 
 \caption{Steady--state values of the first moments of the magnetization $x$ for a system with two communities $M=2$, $N=200$ agents. Right column, unbalanced case: $z_2=0$, $z_1^+=1$, $z_1^-=0$. Left column, balanced case $z_2=0$, $z_1^+=z_1^-=1$ . Different symbols (simulations) and lines (theory) inside each plot correspond to $N_1=10$ (squares), $100$ (circles), and $190$ (triangles).}
 \label{fig:15}
\end{figure}

The latter property makes the system of equations for the moments of degree $D$ to depend only on those of degree less or equal to $D$ and then lead to a complicated but close set of equations for the moments that can be solved exactly and analytically. As a direct application, we show in Fig.~\ref{fig:15} that the first moments of the global and partial magnetizations for the different components of the community are in general different.

\section{Reconstruction of a (discrete) probability distribution from its moments\label{app:2}}

Let $P(x)$ be any probability distribution defined at the discrete set of points $\{x_i=2\frac{i}{N}-1;\,i=0,1,\dots,N\}$, and let $\mathcal F$ be the Hilbert space of normalizable functions $f(x)$ defined in that same set with the scalar product $\langle f,g\rangle=\sum_{i=0}^Nf(x_i)g(x_i)$. $P(x)$ can be expanded in any basis of $\mathcal F$. Amongst all possible basis we choose the orthonormal set of discrete Chebyshev polynomials $\{C_j^N(x)\}_{j=0}^{N}$ \cite{alfe07,mami08,ze75}:
\begin{equation}
 \label{eq:a5}
 \begin{split}
 C_j^N(x)=&\frac{1}{\sqrt{\binom{N+j+1}{2j+1}\binom{2j}{j}}}\sum_{s=0}^{j}(-1)^{s+j} \\ &
 \times \binom{s+j}{s}\binom{N-s}{N-j}\binom{\frac{N}{2}(1+x)}{s},
 \end{split}
\end{equation}
($C_j^N(x)$ is of degree $j$). This basis is obtained from the basis of monomials $\{x^j\}_{j=0}^{N}$ by the Gram-Schmidt orthonormalization procedure. When $N$ tends to infinity, $\mathcal F$ tends to the space of square integrable functions in $[-1,1]$ and $C_j^N(x)$ to the Legendre polynomials \cite{ze75}. The expansion reads:
\begin{equation}
 \label{eq:a6}
 P(x)=\sum_{j=0}^{N} q_jC_j^N(x),
\end{equation}
where $q_j$ are coefficients to be determined. The advantage of using the base of discrete Chebyshev polynomials is that the coefficients $q_j$ can be easily obtained from the orthonormality condition $\medio{C_j^NC_{j'}^N}=\delta_{j,j'}$ as $ q_j=\medio{C_j^N,P}=\sum_i P(x_i)C_j^N(x_i)$, which is nothing but the average value $\medio{C_j^N(x)}$ with respect to the probability $P(x)$. This way, we reconstruct $P(x)$ from its moments, a simple way of solving the so--called moment problem \cite{ak65,mn08} for the present case.

In practice, for large $N$, we can approximate $P(x)$ as
\begin{equation}
\label{eq:a8}
P(x)\approx \sum_{j=0}^{K} \medio{C_j^N(x)}C_j^N(x),
\end{equation}
and express $P(x)$ from the knowledge of the fist moments $\medio{x^k},\,k=1,\dots,K$ of the probability distribution. The approximation turns out to be good if the probability function is not close to zero. Otherwise, the approximation might not respect the important condition $P(x)\ge 0$.

Equation \eqref{eq:a6} is particularly useful with rates at Eq.~\eqref{eq:1} of the general noisy voter model with zealots, since we can analytically compute the moments $\medio{n^j}$ (and $\medio{x^j}$) in the steady state, as we showed in Appendix \ref{app:3}.

\section{Alternative derivation of the critical lines \label{app:1}}
Consider one single community $M=1$ and general rates $\pi^\pm(n)$. From the master equation for the steady probability function $P(n)$ of the system, it is easily inferred the following useful relation
\begin{equation}
 \label{eq:a1}
 P(n)=\frac{\pi^+(n-1)}{\pi^-(n)}P(n-1),
\end{equation}
for $1\le n\le N$. This equation, together with the normalization condition for $P(n)$, provides a closed form for $P$ in terms of the rates
\begin{equation}
 \label{eq:a2}
 P(n)=\frac{\prod_{k=1}^n \frac{\pi^+(n-k)}{\pi^-(n+1-k)}}{1+\sum_{l=1}^N \prod_{k=1}^l \frac{\pi^+(l-k)}{\pi^-(l+1-k)}}.
\end{equation}
Moreover, we can identify the local maximum (minimum) of $P(n)$, and hence infer the shape of $P(n)$, as those values $n_m$ that satisfy
\begin{equation}\label{eq:a3}
P(n_m-1)\le (\ge) P(n_m)\ge (\le) P(n_m+1).
\end{equation}
Expression \eqref{eq:a1} proves that the only possible way $P(n)$ can have more than one extreme is the rates to be nonlinear functions of $n$ (the condition is not sufficient). If the rates are linear, then $P(n)$ has one extreme at most.

If Eqs.~\eqref{eq:a1} and \eqref{eq:a3} are used with general rates of the form \eqref{eq:2}, for the one--community case $M=1$, the extreme of $P(x)$ reads
\begin{equation}
 \label{eq:a4}
 x_m=\frac{N+1}{N}\frac{a_1^+-a_1^-}{a_1^++a_1^--\frac{2h_1}{N}},
\end{equation}
where $x_m=2n_m/N-1$.

Now we particularize Eq.~\eqref{eq:a4} for the rates of the noisy voter model with global influence of zealots, using Eq.~\eqref{eq:2} with $k=1$ for the coefficients. This shows that $x_m=0$ if $z^+=z^-$ (balanced case), and provides Eq.~\eqref{eq:10} for the unbalanced case. Furthermore, from Eq.~\eqref{eq:a4} we also determine the critical value $(a/h)_c$ separating EA and AU phases by imposing $x_m=\pm 1$. The resulting expression for $(a/h)_c$ coincides with Eq.~\eqref{eq:9}, which by the way also provides the critical value for the balanced case. Moreover, for the balanced case, it can be explicitly seen, by imposing Eq.~\eqref{eq:a3} for all $n_c\in\{2,\dots,N\}$, that the uniform solution occurs only when $a/h=(a/h)_c$, with $(a/h)_c$ given by Eq.~\eqref{eq:9}.

The same procedure can be followed by now using Eq.~\eqref{eq:a4} with the effective coefficients obtained in Sec. \ref{sec:4}, namely Eqs. \eqref{eq:11} and \eqref{eq:14}. That way we derive Eqs. \eqref{eq:12} and \eqref{eq:15} for the critical lines.


\begin{thebibliography}{99}
\bibitem{clsu73} P. Clifford and A. Sudbury, Biometrika \textbf{60} 581-588 (1973)
\bibitem{holi75} R. A. Holley and T. M. Liggett, Ann. Probab. \textbf{3} 643-663 (1975)
\bibitem{li05} T. M. Liggett. \emph{Interacting Particle Systems}, Springer, 2005 Reprint (Classics in Mathematics)
\bibitem{castellano09} C. Castellano, S. Fortunato, and V. Loreto, Rev. Mod. Phys. \textbf{81},
591 (2009).
\bibitem{kaprisky10}P. L. Kaprisky, S. Redner and E. Ben Naim \emph{A kinetic view of Statistical Physics}, Cambridge University Press, 2010.
\bibitem{pejorawa17} M. Perc, J. J. Jordan, D. G. Rand, Z. Wang, S. Boccaletti, and A. Szolnoki, Physics Reports \textbf{687} 1-51 (2017)
\bibitem{suchecki}K. Suchecki, V. M. Eguiluz,  and M. San Miguel, Physical Review E \textbf{72}, 036132 (2005).
\bibitem{mo03} M. Mobilia, Phys. Rev. Lett. \textbf{91} 028701 (2003)
\bibitem{mopere07} M. Mobilia, A. Petersen, and S. Redner, J. Stat. Mech. \textbf{08} P08029 (2007)
\bibitem{fues14} M. I. D. Fudolig and J. P. H. Esguerra, Physica A \textbf{413} 626-634 (2014)
\bibitem{X15} D.D. Chinellato, I. R. Epstein, D. B. Y. Bar-Yam, and M. A. M. de Aguiar, J. Stat. Phys. \textbf{159} 221-230 (2015)
\bibitem{kahoda14} H. Kashisaz, S. S. Hosseini, and A. H. Darooneh, Physica A \textbf{402} 49-57 (2014)
\bibitem{moscag15} C. A. Moreira, D. M. Schneider, and M. A. M. de Aguiar, Phys. Rev. E \textbf{92} 042812 (2015)
\bibitem{mo15} M. Mobilia, Phys. Rev. E \textbf{92} 012803 (2015)
\bibitem{memozi16} A. Mellor, M. Mobilia, and R. K. P. Zia, Europhys. Lett. \textbf{113} 48001 (2016)
\bibitem{memozi17} A. Mellor, M. Mobilia, and R. K. P. Zia, Phys. Rev. E \textbf{95} 012104 (2017)
\bibitem{gamo91} S. Galam and S. Moscovici, Eur. J. of Soc. Psych. \textbf{21} 49 (1991)
\bibitem{ga97} S. Galam, Physica A \textbf{238} 66 (1997)
\bibitem{gaja07} S. Galam and F. Jacobs, Physica A \textbf{381} 366-376 (2007)
\bibitem{xisrkozhlisz11} J. Xie, S. Sreenivasan, G. Korniss, W. Zhang, C. Lim, and B. K. Szymanski, Phys. Rev. E \textbf{84}, 011130 (2011)
\bibitem{ma12} N. Masuda, Scientific Reports \textbf{2} 646 (2012)
\bibitem{librwashstha13} Q. Li, L. A. Braunstein, H. Wang, J. Shao, H. E. Stanley, S. Havlin, J. Stat. Phys. \textbf{151} 1 (2013)
\bibitem{ma13} N. Masuda, Phys. Rev. E \textbf{88} 052803 (2013)
\bibitem{tuwegr13} M. Turalska, B. J. West, P. Grigolini, Scientific Reports \textbf{3} 1371 (2013)
\bibitem{veswch14} G. Verma, A. Swami, and K. Chan, Physica A \textbf{395} 310--331 (2014)
\bibitem{nama15} Y. Nakajima and N. Masuda, J. Math. Biol. \textbf{70} 465-484 (2015)
\bibitem{coca16} F. Colaiori and C. Castellano, J. Stat. Mech. \textbf{03} 033401 (2016)
\bibitem{klwidodo17} P. P. Klamser, M. Wiedermann, J. F. Donges, and R. V. Donner, ArXiv:1612.06644
\bibitem{szpe16} A. Szolnoki and M. Perc, Phys. Rev. E \textbf{93} 062307 (2016)
\bibitem{grma95} B. L. Granovsky and N. Madras, The Noisy Voter Model. Stochastic Processes and their Applications \textbf{55} 23-43 (1995)
\bibitem{ki93} A. Kirman, Q. J. Econ. \textbf{108}, 137-156 (1993).
\bibitem{alluwa05} S. Alfarano, T. Lux, and F. Wagner, Computational Economics \textbf{26} 19-49 (2005)
\bibitem{alluwa08} S. Alfarano, T. Lux, and F. Wagner, Journal of Economic Dynamics and Control \textbf{32} 101-136 (2008)
\bibitem{almi09} S. Alfarano and M. Milakovic, Journal of Economic Dynamics and Control \textbf{33} 78-92 (2009)
\bibitem{almira13} S. Alfarano, M Milakovic, and M Raddant, The European Journal of Finance \textbf{19} 449-465 (2013)
\bibitem{deneamwe00} G. Deffuant, D. Neau, F. Amblard, and G. Weisbuch, Advances in Complex Systems \textbf{3} 87-98 (2000)
\bibitem{wedeamna02} G. Weisbuch, G. Deffuant, F. Amblard, J. P. and Nadal, Complexity \textbf{7}(3) 55-63 (2002) 
\bibitem{hekr02} R. Hegselmann and U. Krause, Journal of Artificial Societies and Social Simulation \textbf{5} 1-24 (2002)
\bibitem{wedeamna03} G. Weisbuch, G. Deffuant, F. Amblard, and J. P. Nadal, \emph{Heterogenous Agents, Interactions and Economic Performance, chapter Interacting Agents and Continuous Opinions Dynamics}, 225-242. Springer Berlin Heidelberg (2003) 
\bibitem{catosa15} A. Carro, R. Toral, M. San Miguel, PLoS ONE 10(7): e0133287 (2015)
\bibitem{catosa16} A. Carro, R. Toral, M. San Miguel, Sci. Rep. \textbf{6} 24775 (2016)
\bibitem{lato13} F. L. Lafuerza and R. Toral, Sci. Rep. \textbf{3} 1189 (2013)
\bibitem{fesurasaeg14} J. Fern\'andez-Gracia, K. Suchecki, J. J. Ramasco, M. San Miguel, and V. Egu\'{\i}luz, Phys. Rev. Lett. \textbf{112} 158701 (2014)
\bibitem{alfe07} A. Eisinberg and G. Fedele, International Mathematical Forum \textbf{2} 21 1007-1020 (2007)
\bibitem{mami08} G. Mastroianni, G.V. Milovanovic, Interpolation Processes. Basic Theory and Applications, Springer-Verlag (2008)
\bibitem{ze75} G. S. Zego, Orthogonal Polynomials, Amer. Math. Soc. Colloq. Publ. XXIII, Amer. Math. Soc. (1975)
\bibitem{ak65} N. I. Akhiezer, The Classical Moment Problem and Some Related Questions in Analysis, Oliver and Boyd (1965)
\bibitem{mn08} R. M. Mnatsakanov, Statistics and Probability Letters \textbf{78} 1869-1877 (2008)
\bibitem{crblan14} N. Crokidakis, V. H. Blanco, and C. Anteneodo, Phys. Rev. E \textbf{89} 013310 (2014)
\end{thebibliography}
\end{document}